# Multi-scale seismic envelope inversion using a direct envelope Fréchet derivative for strong-nonlinear full waveform inversion

*Ru-Shan Wu, Guo-Xin Chen*[1]
*Modeling and Imaging Laboratory, Earth & Planetary Sciences, University of California, Santa Cruz, USA*

## Abstract

Traditional seismic envelope inversion takes use of a nonlinear misfit functional which relates the envelope of seismogram to the observed wavefield records, and then derive the sensitivity kernel of envelope to velocity through the use of waveform Fréchet derivative by linearizing the nonlinear functional. We know that the waveform Fréchet derivative is based on the Born approximation of wave scattering and can only work for the case of weak scattering. Therefore, the traditional envelope inversion using waveform Fréchet derivative has severe limitation in the case of strong-scattering for large-scale, strong-contrast inclusions. In this paper, we derive a new *direct envelope Fréchet derivative* (sensitivity operator) based on energy scattering physics without using the chain rule of differentiation. This new envelope Fréchet derivative does not have the weak scattering assumption for the wavefield and can be applied to the case of strong-nonlinear inversion involving salt structures. We also extend the envelope data into multi-scale window-averaged envelope (WAE), which contains much rich low-frequency components corresponding to the long-wavelength velocity structure of the media. Then we develop a joint inversion which combines the new multi-scale direct envelope inversion (MS-DEI) with standard FWI to recover large-scale strong-contrast velocity structure of complex models. Finally, we show some numerical examples of its successful application to the inversion of a 1-D thick salt-layer model and the 2D SEG/EAGE salt model.

## 1 Introduction

It is well-known that the conventional full waveform inversion (FWI) is based on the weak scattering theory and can only be applied to the weak-nonlinear full waveform inversion. In the classic papers (Tarantola, 1984a; Laily, 1983; Tarantola, 1984b) which introduced the Newton-type local optimization into modern seismic inversion, it is clearly stated that the theory and method are based on the weak-scattering theory and can only apply to the weak nonlinear inversion. One severe consequence of the weak-scattering assumption is the initial model dependence of FWI. Since parameter perturbations have to be small to satisfy the weak-scattering assumption, therefore, the starting model has to be close to the true model. This requirement appeared become a impassable roadblock for FWI to apply to the case of large-scale strong-contrast inclusions, such as irregular salt domes.

In order to reduce the nonlinearity of full-waveform inversion, investigators introduced different kinds of nonlinear data functional with the goal that those functionals will have better linearity to velocity perturbation than the waveform data, The nonlinear data functional used in the literature ranges from traveltime (Luo and Schuster, 1991; Tromp et al., 2005; Fitchner et al., 2008: Choi and Alkhalifah, 2013; Luo et al., 2016) to envelope (Bozdag, 2011; Wu et al., 2012, 2013, 2014; Chi et al., 2014; Luo et al., 2015). However, in order to used the convenient and mature tool of gradient method based on the theory of waveform Fréchet derivative, as a routine the nonlinear functional is linearized and the chain rule of differentiation is used to relate the Fréchet derivative of the new functional (sensitivity operator of the new functional) to the familiar waveform Fréchet derivative. This procedure makes the inversion using new nonlinear data functional fall back to the local minima trap of traditional FWI. Although the new functional may offer certain advantages, but these advantages are severely reduced and limited. In this paper, we will concentrate on solve this problem for seismic envelope inversion and will not discuss other nonlinear functionals. As we mentioned above, seismic envelope inversion has recently been introduced and developed to recover the long-wavelength background structure when seismic source lacks ultra-low frequency source (below 5Hz or so). It has been pointed out that the envelope inversion involves a nonlinear scale separation (through the envelope operator or the demodulation operator) to extract the ultra-low frequency signal which is the modulation signal riding on the carrier signal. However, when applied to the envelope-inversion for large structures of strong-scattering media, such as those involved with salt, basalt or karst, the success is very limited due to the linearization of the nonlinear functional and the use of the waveform Fréchet derivative as we mentioned above. This approximation renders the current envelope inversion basically a weak-nonlinear approximation and therefore is not applicable in the strong-nonlinear case.

[1] Visiting from School of Earth Sciences, Zhejiang University, China

In addition to the problem introduced by the linearization, the current seismic envelope inversion has another problem when applied to strong-nonlinear waveform inversion. The instantaneous amplitude curves extracted by analytical signal transform using the Hilbert transform are highly fluctuating curves and may still cause cycle-skipping during seismic envelope inversion, even though the cycle-skipping is much weaker than the traditional FWI.

In this paper, we propose a multi-scale direct envelope inversion (MS-DEI) and derive a new direct envelope Fréchet derivative (EFD) without relying on waveform Fréchet derivative for implementation. With a joint misfit functional of multi-scale envelope and waveform the new method can be applied to strong-nonlinear inversions. Numerical tests on a simple salt layer model and the SEG 2D salt model demonstrate the validity of the approach.

## 2. New Direct Envelope Fréchet derivative for multi-scale envelope inversion

### 2.1 Problem of using traditional waveform Fréchet derivative

Traditionally, envelope Fréchet derivative (envelope sensitivity kernel) is derived from a chain rule of functional derivatives, and the implementation is realized by the use of waveform Fréchet derivative. Define data functional as the window-averaged envelope (WAE) $e_W(t)$ as

$$e_W(t) = d(u(t),t) = \frac{1}{\tau_W}\int dt' W(t-t')[u^2(t') + u_H^2(t')] \quad (1)$$

where $u(t)$ is the seismic wavefield, i.e. the seismic waveform data (seismograms), which could be pressure of particle-velocity records, and $u_H$ is the corresponding Hilbert transformed records, $d(u(t),t)$ is the data functional, here is defined as envelope which is a functional of waveform data. $W(t)$ is a window function and $\tau_W$ is its effective width. So the data residual becomes

$$r_W(t) = e_{W,syn}(t) - e_{W,obs}(t) \tag{2}$$

The misfit functional is defined as

$$\sigma(r) = \frac{1}{2}\sum_{S,R}\int_0^T \left(r_W(t) r_W(t)\right) dt \tag{3}$$

where the summation is over the sources, receivers, and recording time, and the data residual $r_W(t)$ in (2) can be written explicitly

$$r_W(t) = \frac{1}{\tau_W}\int dt' \mathrm{W}(t-t')[y^2(t') + y_H^2(t') - u^2(t') - u_H^2(t')] \tag{4}$$

where $y(t)$ is the synthetic record and $u(t)$ is the observed record. Conventional envelope inversion derived the gradient by linearization of the local minimization problem using the chain rule of differentiation (Bozdag *et al.*, 2011; Wu *et al.*, 2014; Chi *et al.*, 2014; Luo & Wu, 2015; Bharadwaj *et al.*, 2016),

$$\begin{aligned}\frac{\partial\sigma}{\partial v} &= \sum_{S,R}\int_0^T \frac{\partial e_W(t)}{\partial v} r_W(t) dt \\ &= \sum_{S,R}\int_0^T \left[\frac{\partial e_W(t)}{\partial y}\frac{\partial y}{\partial v} + \frac{\partial e_W(t)}{\partial y_H}\frac{\partial y_H}{\partial v}\right] r_W(t) dt \\ &= \sum_{S,R}\int_0^T dt r_W(t) \frac{2}{\tau_w}\int dt' W(t-t')[y(t')\frac{\partial y}{\partial v} + y_H(t')\frac{\partial y_H}{\partial v}]\end{aligned} \tag{5}$$

where v(x) is the velocity distribution, $\frac{\partial y}{\partial v}$ and $\frac{\partial y_H}{\partial v}$ are the *Fréchet derivatives for the waveform data*.

Note that in the first row of above equation, in operator form it can be written as

$$\frac{\partial\boldsymbol{\sigma}}{\partial\mathbf{v}} = \left(\frac{\partial\mathbf{e}_W}{\partial\mathbf{v}}\right)^T \mathbf{r}_W \tag{6}$$

where $\mathbf{r}_W$ is the MS envelope data residual (a vector), and $\partial\mathbf{e}_W/\partial\mathbf{v}$ in fact is the Fréchet derivative operator for the window-averaged envelope data with respect to velocity perturbation, $(\ )^T$ denotes the *transpose operator* or the approximate *adjoint operator*, which involves spatial-transpose and time-reversal (backpropagation) of the operator (see Tarantola, 2005, section 7.18). In order to use the waveform Fréchet derivative (WFD) for implementation, traditionally the chain rule of differentiation is applied to derive the relation between the envelope Fréchet derivative (EFD) and the waveform Fréchet derivative as the following

$$\mathbf{F}_{env} = \frac{\partial\mathbf{e}_W}{\partial\mathbf{v}} = \frac{\partial\mathbf{e}_W}{\partial\mathbf{y}}\frac{\partial\mathbf{y}}{\partial\mathbf{v}} = \frac{\partial\mathbf{e}_W}{\partial\mathbf{y}}\mathbf{F}_{wav} \tag{7}$$

where $\mathbf{F}_{env} = \partial\mathbf{e}_W/\partial\mathbf{v}$ is the Fréchet derivative of envelope to velocity, $\partial\mathbf{e}_W/\partial\mathbf{y}$ is the envelope functional derivative of envelope to waveform data, which is a linearization of envelope-waveform functional, and $\mathbf{F}_{wav} = \partial\mathbf{y}/\partial\mathbf{v}$ is the waveform Fréchet derivative with respect to velocity, a linearization of waveform-velocity functional (see appendices A and B). It is known that the relation between different sensitivity operators in the above derivation is based on the weakly nonlinear assumption, and is not valid for *strongly nonlinear data functional*s (For detailed derivation and physical reasoning, see Appendix A and B. In appendix A, the nonlinearity of the waveform functional is summarized based on Wu and Zheng, 2014. In appendix B the severe limitation of applying the chain rule of differentiation to stronger nonlinear functional is discussed). For conventional FWI using scattered wavefield as data, the data functional (waveform data as functional of model variation) can be nearly linear or weak-nonlinear if the background velocity structure is close to the true structure; However, if the starting model is far from the true model such as shown in our numerical tests for the SEG salt model in Figure 6 (the starting model is a 1-D linear model) , the observed scattered wavefield will be so strong and drastically different from the wavefield produced by the initial model which in this case cannot generate any reflection, so linearization has no physical basis and becomes meaningless. To further elucidate the situation, we draw some equations from Appendix B to illustrate the limitation of linearization applying to the derivation of indirect sensitivity operators. For general nonlinear sensitivity operators, the chain rule (7) can be generalized to (see equation (B7))

$$\delta \mathbf{e} = \left(\frac{\delta \mathbf{e}}{\delta \mathbf{u}}\right)_{NL} \left(\frac{\delta \mathbf{u}}{\delta \mathbf{v}}\right)_{NL} \delta \mathbf{v} \qquad (8)$$

where $(.)_{NL}$ means the sensitivity operator is nonlinear. Expandxpand both nonlinear operators into Taylor series and substitute back into above equation, resulting in (equation (B10)

$$\left(\frac{\delta \mathbf{e}}{\delta \mathbf{v}}\right)_{NL} = \frac{\delta \mathbf{e}}{\delta \mathbf{u}}\frac{\delta \mathbf{u}}{\delta \mathbf{v}} + \frac{1}{2!}\left[\frac{\delta \mathbf{e}}{\delta \mathbf{u}}\frac{\delta^2 \mathbf{u}}{\delta \mathbf{v}^2}\delta \mathbf{v} + \frac{\delta^2 \mathbf{e}}{\delta \mathbf{u}^2}\delta \mathbf{u}\frac{\delta \mathbf{u}}{\delta \mathbf{v}}\right] + \cdots\cdots. \qquad (9)$$

Neglecting all the higher-order terms in above equation (linearization) results in the standard linearized equation (7). We see that the conventional approximation (7) is a ***double linearization*** which is only valid for case of weak nonlinearity. The higher order terms are model-dependent and drop of all the higher order terms will lose the low-frequency information in the sensitivity operator.

### 2.2 **Direct envelope Fréchet derivative (DEFD) for strong-nonlinear data functional**

As we discussed in previous section, in the case of strong nonlinearity, the chain rule of differentiation (linearization) is nolonger applicable. Derivation based on the linearization resulted in the severe limitation of the envelope inversion applying only to weak scattering case, such as the Marmousi model. To our best knowledge It has never been successfully applied to strong-nonlinear cases such as the SEG 2D salt model or other models having large salt structures with a 1-D linear initial model. In this paper, we try to derive the direct envelope Fréchet derivative without using the chain rule. In fact envelope formation can be formulated based on the theory of energy scattering (see Wu *et al*., 2016). Due to the additivity by neglecting the interference in energy scattering, linear superposition is valid under the single scattering approximation for velocity (or impedance) perturbation, leading to a better linearity in the case of strong scattering such as the boundary scattering (reflection) of strong-contrast media. This is why we prefer to derive the *envelope Fréchet derivative* (EFD) directly with energy formulation.

2.2.1 Sensitivity operator and virtual source operator.

To facilitate the derivation, we first reformulate the waveform inversion in an operator form. In waveform inversion, the *sensitivity operator* (Fréchet derivative) can be expressed in an operator form (Tarantola, 1987, 2005; Pratt *et al*, 1998)

$$\delta\mathbf{u} = \mathbf{F}_u\delta\mathbf{v} = \mathbf{G_0 Q_0}\delta\mathbf{v} \tag{10}$$

where $\delta\mathbf{u}$ is the wavefield change due the velocity perturbation $\delta\mathbf{v}$, $\mathbf{F}_u$ is the Fréchet Fréchet derivative of the waveform data (similar to the Jacobien matrix **J** in Pratt *et al*, 1998), $\mathbf{G_0}$ is the background Green's operator, and $\mathbf{Q_0}$ is the linearized *virtual source operator*, defined as (under weak scattering approximation)

$$Q_0(x,x',t) = -\frac{2}{v_0^3(x')}\frac{\partial^2}{\partial t^2}u_{\mathbf{0}}(x',t)\delta\left(x-x'\right) \tag{11}$$

where $v_0(x)$ is the background velocity and $u_0(x,t) = g_0(x,t;x_s)$ is the local incident wavefield excited by a shot at $x_s$. We see that in the case of scalar wave equation, the virtual source operator is a diagonal operator and therefore is called *virtual source term* (Pratt *et al*., 1998).

2.2.2 Virtual source operator(VSO) for envelope inversion

Similar to (10) for the waveform inversion, we can write the envelope sensitivity operator in an operator form

$$\delta\mathbf{e} = \mathbf{F}_E\delta\mathbf{v} = \mathbf{G}_E\mathbf{Q}_E\delta\mathbf{v} \tag{12}$$

Where $\delta\mathbf{e}$ is the envelope variation due to velocity perturbation $\delta\mathbf{v}$, $\mathbf{F}_E$ is the envelope Fréchet (EFD), $\mathbf{G}_E = \mathbf{G}_0^{(e)}$ is the envelope Green's operator, and $\mathbf{Q}_E = \mathbf{Q}_0^{(e)}$ is the envelope virtual source operator. We formulate the envelope modeling and inversion based on energy scattering theory. In Appendix C we derive the energy scattering formulas for both the weak scattering (Born scattering) and strong boundary scattering (boundary reflections).

For envelope inversion, virtual source operators for strong scattering, such as boundary reflection, is very different from the weak scattering, such as weakly perturbed volume heterogeneity. In this paper, we use only the information contained in the instantaneous amplitude (or "envelope amplitude"), and neglect the information carried by the instantaneous phase. Therefore the physics of envelope amplitude can be modeled by energy scattering. The case of weak volume scattering is similar to the Born approximation for energy scattering. Then we have an envelope (energy) virtual source operator (VSO) as

$$Q_v^{(e)}(x,x',t) = |\frac{4}{v_0^6(\mathbf{x})}\frac{\partial^4}{\partial t^4}\left|G_0\right|^2(x',t;x_s)\delta\left(x-x'\right) \tag{13}$$

where "(e)" for superscript denotes for "energy", and $|G_0|^2 = |u_0|^2$ is the energy Green's function, here the incident energy pulse from the source. The VSO relates the envelope energy residual $\delta\mathbf{e}^2$ and the squared velocity perturbation $(\delta v)^2$ through

$$\delta\mathbf{e}^2 = |\delta\mathbf{u}|^2 = \left|\mathbf{G_0}\right|^2 \mathbf{Q}_v^{(e)} (\delta\mathbf{v})^2 \tag{14}$$

The detailed derivation based on energy scattering theory is in Appendix C. We see from the derivation that for energy scattering, the time-differentiation becomes fourth order, and the magnitude of the operator is inversely proportional to the six power of the background velocity. This is consistent with case of wave scattering, since scattered energy is proportional to the squared amplitude of wavefield. In general, the energy scattering inversion for weak scattering is similar to FWI but the forward and inverse processes are all done through energy instead of wavefield. So many limitations imposed by the weak scattering assumption remain unsurpassable. In this paper we mainly concentrate on the strong scattering case, and will not go into details for the weak scattering case.

### 2.2.3 Virtual source operator(VSO) for strong scattering (boundary scattering)

For the case of boundary scattering (backscattering or reflection) of strong-contrast media, wave and energy scattering can be quite different from weak scattering. Boundary reflection is formed by coherent scattering by large volume strong-contrast inclusions (such as salt bodies, carbonate rocks, etc.). For a large salt body, the mutual interference between scattered waves from different volume elements leads to the total cancellation of scattered waves from the salt interior and the formation of strong reflections from the boundaries and the transmitted wave penetrating the volume. This is why boundary reflection is treated totally different mathematically from volume heterogeneities. Wave equations are set for inner and outer regions and boundary conditions are to be matched along boundaries. Strong coherent interaction between scattered waves is the root of strong nonlinearity in FWI. In this paper we apply energy scattering directly to boundary scattering to mitigate the strong nonlinearity. Based on a surface representation integral (Kirchoff integral) of the boundary scattered waves, we derive an energy scattering formulation and its corresponding virtual source operator ((C33) and (C39) in Appendix C). From the derivation, we write the scattered energy by boundaries of strong-contrast as (C41)

$$\delta\mathbf{e}^2 = |\delta\mathbf{u}|^2 = \left|\mathbf{G}_F\right|^2 \left|\mathbf{Q}_\gamma\right|^2 \boldsymbol{\gamma}^2 \tag{15}$$

$\delta\mathbf{e}^2$ is the scattered energy (squared envelope) data residual, $\mathbf{G}_F$ is forward-scattering renormalized Green's operator (for homogeneous media, just the free-space Green's operator), $\left|\mathbf{Q}_\gamma\right|^2$ is the energy virtual source operator for <u>energy reflection coefficient</u> $\boldsymbol{\gamma}^2$. The kernel of $\left|\mathbf{Q}_\gamma\right|^2$ is

$$\left|Q_\gamma\right|^2 (x, x', t) \doteq 4 \left|G_F\right|^2 (x', t; x_s) \delta\left(x - x'\right) \tag{16}$$

In appendix C we show why the virtual source operator of boundary scattering is different from the volume scattering. In physical reasoning, we see the different frequency-dependences of weak scattering (Born scattering) and strong scattering (boundary scattering): Born scattering has a frequency-square dependence, while boundary reflection is frequency-independent. Some authors tried to use the renormalization theory to build a bridge connecting boundary reflection and volume scattering (e.g. Kirkinis, 2008; Wu and Zheng, 2014; Wu et al., 2016). Wavefield renormalization happens due to strong constructive and destructive interferences. It modifies nonlinearly the weak scattering in two ways. One is the elemination of scattered field from interior volume elements, so only reflected and transmitted waves from the boundaries exist; The other is the change of frequency-dependence (f-square) into frequency-independence. These can be explained physically by wave interference (renormalization process). The reflection is formed by constructive and destructive interferences. The constructive one is mainly coming from the Fresnel zone near the surface, from which the scattered waves are basically in-phase so they reinforce with each other. Beyond this zone, the scattered waves cancel each other due to destructive interference, so there is no contribution beyond this boundary layer. We know that low-frequency waves have large Fresnel zone, so the contributing volume (the integration volume) will be larger so that their total contribution will be the same as the high-frequency waves. Due to the f-independence, boundary reflection is a wide-band response to the incident energy-pulse, while Born backscattering is a high-pass filtering process. The use of Born-type virtual source operator on reflected waves will filter out the low-frequency information in the envelope residual, leading to difficulty in recovering long-wavelength velocity structures.  In this paper, we deal with boundary scattering of strong contrast inclusions such as salt bodies, therefore, the virtual source operator for strong scattering defined in (C39) and (16) is adopted.

From equation (C38) in appendix C, we have energy reflection coefficient and reflection coefficient  under small-angle approximation for the case of constant density written as

$$
\begin{aligned}
\gamma_e &= \gamma^2 \approx \frac{(v_2 - v_1)^2}{(v_2 + v_1)^2} = \frac{(\Delta v)^2}{(2v_0 + \Delta v)^2}, \\
\gamma &= sign(\gamma)\sqrt{\gamma_e} \approx \frac{\Delta v}{2v_0 + \Delta v}, \\
|\gamma| &= \frac{|\Delta v|}{2v_0 + \Delta v}
\end{aligned}
\tag{17}
$$

where $v_0$ is the background velocity field (smooth long-wavelength structure) without salt bodies, $\Delta v$ is the *velocity contrast* between the inclusions and the background. Note that $\Delta v$, in the case with salt inclusions, could be larger than $v_0$, so weak perturbation method is not appropriate. That is why our method is based on boundary reflection for strong scattering case. Also we note that in the current version of direct envelope inversion based on energy scattering, we did not use the polarity information of reflection signal, and therefore the sign of the reflection coefficient cannot be recovered. As a remedy, we apply a joint misfit functional with both the envelope data and the waveform data to take care the sign of velocity update.

Based on this approximation, we may change the sensitivity operator from energy reflection coefficient to velocity perturbation,

$$\delta \mathbf{e}^2 = |\delta \mathbf{u}|^2 = \left|\mathbf{G}_F\right|^2 \left|\mathbf{Q}_v\right|^2 |\delta \mathbf{v}|^2 = \mathbf{F}_E \delta \mathbf{v}^2 \tag{18}$$

Here $\delta \mathbf{e}^2$ as a vector is defined as $\delta \mathbf{e}^2 = [\delta e_j^* \delta e_j]$. The kernel of VSO $\left|\mathbf{Q}_v\right|^2$ is

$$\left|Q_v\right|^2 (x, x', t) \doteq \frac{4}{(2v_0 + \Delta v)^2} \left|G_F\right|^2 (x', t; x_s) \delta\left(x - x'\right) \tag{19}$$

where $\mathbf{F}_E$ is the Fréchet derivative of energy data (square of envelope ampletude) to squared velocity perturbations. In equation (18) and (19), we derive the envelope sensitivity operator (Fréchet derivative) directly based on energy scattering theory, and therefore no weak scattering assumption or weak nonlinearity of waveform sensitivity operator is imposed. This ***direct envelope Fréchet derivative*** can improve the convergence of envelope inversion for strong scattering case and is critical for long-wavelength recovery in multi-scale direct envelope inversion.

### 2.3 **Adjoint operator method for multi-scale envelope inversion using the new direct Fréchet derivative**

Defining envelopes as data, i.e. $\mathbf{d} \triangleq \mathbf{e}$, then we can apply the adjoint operator of $\mathbf{F}_E$ to derive the gradient operator for envelope inversion. Apply $\mathbf{F}_E^T$, which is the transpose (loosely speaking: the adjoint) of $\mathbf{F}_E$, to equation (18) , resulting in

$$\delta \mathbf{v}^2 = (\mathbf{F}_E^T \mathbf{F}_E)^{-1} \mathbf{F}_E^T \delta \mathbf{e}^2 \ . \tag{20}$$

This is a generalized linear inversion, and $\mathbf{F}_E^T \mathbf{F}_E$ is recognized as the approximate Hessian operator. Since the operator does not involve the second order derivative, it is in fact an illumination operator, and its inverse is an illumination correction or acquisition aperture correction (see, Wu *et al*., 2004; Yan *et al*., 2014). For the *gradient method*, we need only the adjoint envelope Fréchet:

$$\mathbf{F}_E^T = (\mathbf{G}_E \mathbf{Q}_E)^T = \mathbf{Q}_E^T \mathbf{G}_E^T \ . \tag{21}$$

where $\mathbf{G}_E = \left|\mathbf{G}_F\right|^2$ and $\mathbf{Q}_E = \left|\mathbf{Q}_v\right|^2$. We see that because of the linear relationship between scattered energy $\delta \mathbf{e}^2$ and the squared velocity jumps (or variations) $\delta \mathbf{v}^2$. Then we can establish a weak-nonlinear optimization using Newton's method to invert $\delta \mathbf{v}^2$ from $\delta \mathbf{e}^2$ and then convert to the velocity jumps $\Delta \mathbf{v}$ .

From the above equation, we see that the calculation of envelope Fréchet is efficient by using $\mathbf{G}_E^T$ or $\mathbf{G}_F^T$, which is the transposed (including reverse-time) envelope propagator (backpropagator). In comparison, the *traditional adjoint envelope Fréchet derivative* can be derived from (7)

$$\mathbf{F}_{env}^T = \mathbf{F}_{wav}^T \left( \frac{\partial \mathbf{e}_W}{\partial \mathbf{y}} \right)^T \tag{22}$$

where the linearized approximation of the functional $\left(\partial \mathbf{e}_W / \partial \mathbf{y}\right)^T$ serves as the adjoint source operator applying to the envelope data residual. The severe filtering effect of $\left(\partial \mathbf{e}_W / \partial \mathbf{y}\right)^T$ causes the loss of important information in the corresponding adjoint source.The calculation of $\mathbf{Q}_E^T$ is straightforward. However, the energy or amplitude propagator (Green's function) and backpropagator need to be discussed.

2.3.1 Removal of near-zero frequencies

The purpose of envelope inversion using the new Fréchet derivative is to recover the long-wavelength background velocity structure. The initial model prefered is a 1-D gradient medium or other smooth media without a priori knowledge of the salt body. In these smooth media, the parabolic wave equation will be best for propagation. However, we can also use the wave equation based propagator for this purpose. The energy or envelope propagator is a broad-band pulse propagator, and the backpropagator is an energy focusing operator. The energy or envelope pulses contain rich ULF information, including the zero-frequency. In smooth media as our case for direct envelope inversion, standard wave equation based propagators such as the full-wave finite difference method, or wide-angle one-way propagators will do a good job. Nevertheless, there is a minor problem for the envelope propagator, that is the artifacts caused by the nearfield produced by zero and near-zero frequencies. Nearfield means the field close to the source with distance much smaller than the wavelength. In the case of zero and near-zero frequencies, the region of the corresponding near-field could be huge so the shallow structures may be masked by the near-field artifacts of sources. For this reason, we need to remove the zero and near-zero frequencies from the data or source wavelet to avoid the artifacts. In the meanwhile, the useful ULF components of the multi-scale envelope data, which are in the range of 1-5 Hz in the salt structure recovery cases, of course need to be preserved. Therefore, the cut-off frequency for the near-zero frequency filtering depends on the model size and the scale of salt domes and should be set to be below 1 Hz in our case.

2.3.2 Gradient operator using the direct envelope Fréchet derivative

In operator form, the gradient of the misfit function using the new EFD can be written as (see Pratt *et al.*, 1998)

$$\frac{\delta \boldsymbol{\sigma}}{\delta \mathbf{v}^2} = \mathbf{F}_E^T \delta \mathbf{e}^2 ; \quad \text{or} \quad \frac{\partial \boldsymbol{\sigma}}{\partial |\mathbf{v}|} = \mathbf{F}_e^T \delta \mathbf{e} \tag{23}$$

Where $\delta \mathbf{e}$ is the envelope data residual and $\delta \mathbf{e}^2$ is the envelope energy data residual. For multi-scale envelope inversion, above equation becomes

$$\frac{\partial \boldsymbol{\sigma}}{\partial \mathbf{v}^2} = \mathbf{F}_E^T \delta \mathbf{e}_W = \mathbf{F}_E^T \mathbf{r}_W = \mathbf{Q}_E^T \mathbf{G}_E^T \mathbf{r}_W \tag{24}$$

where $\mathbf{r}_W$ is the multi-scale envelope energy data residual as explicitly expressed in (4). We see that with the new EFD, the envelope data residual is directly back-propagated and maped to the velocity updates. In this way, the *ultra-low-frequency* components in the multi-scale envelope data will be better preserved for the low-wavenumber component recovery. We can also define the illumination-corrected gradient as

$$\hat{\nabla}_v \sigma = (\mathbf{F}_E^T \mathbf{F}_E)^{-1} \frac{\partial \boldsymbol{\sigma}}{\partial \mathbf{v}^2} = (\mathbf{F}_E^T \mathbf{F}_E)^{-1} \mathbf{F}_E^T \delta \mathbf{e}_W \tag{25}$$

Illumination correction will take care the influence of geometric spreading and effective aperture to the gradient field, so the resulted gradient field for deep target will be much improved.

### 2.3.3 Comparison of new data and new adjoint sources with the traditional ones

In this paper, we adopt the window-averaging approach (Wu *et al.*, 2016, Wu & Chen, 2017a, b; Chen *et al.*,2018a,b) to to perform scale-decomposition to the ULF (ultra-low frequency) components in the envelope data which are extracted by a nonlinear envelope operator from seismic records. As we defined in (1), $e_W(t)$ is window-averaged envelope (WAE) which can be seen as a low-pass filtered envelope of the original Hilbert envelope (here we refer to the envelope envergy data). Different windows provide different low-pass bands for the multi-scale decomposition of envelope data. This multi-band filtering of envelope data is very different from the multi-scale decomposition for waveform data, since envelope filtering is after the nonlinear operation of envelope operator. Envelope contains ULF information below the source frequency-band. Figure 1 gives some examples of the envelope curve with different window widths from 400ms to 1000ms for the data of SEG salt model. We can see their differences in the apparent pulse widths. In Figure 2 we compare the corresponding spectra for the case of full-band source (Figure 2a) and low-cut source (from 5Hz below) (Figure 2b). It can be clearly seen that the low-passed envelope data posses the ULF information beyond the lowest frequency of the source spectrum, especially in the low-cut source case.

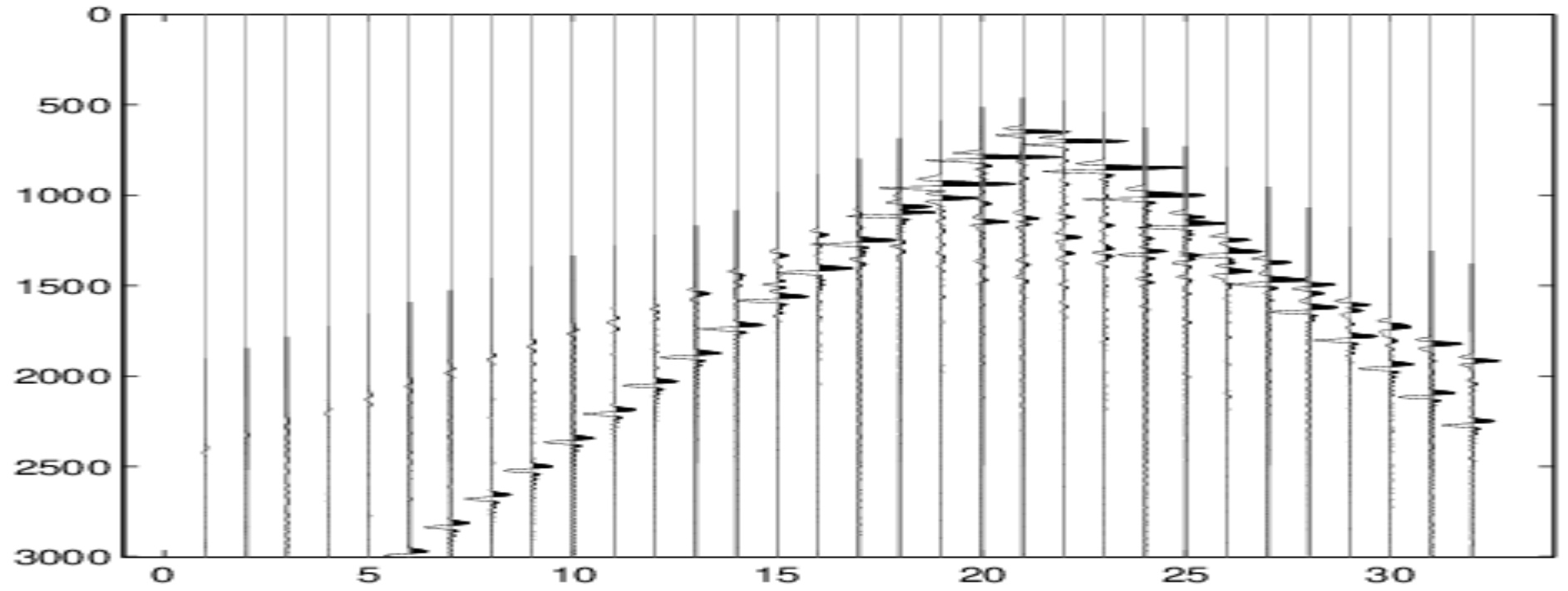

(a)

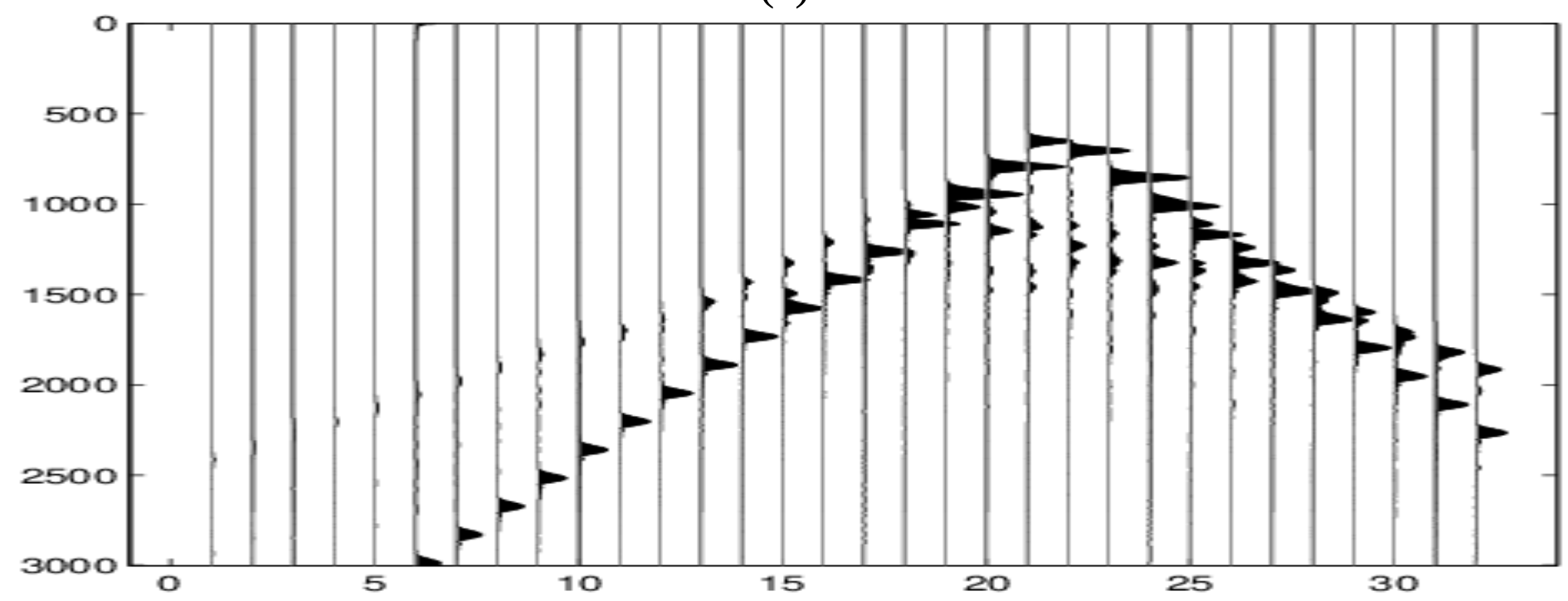

(b)

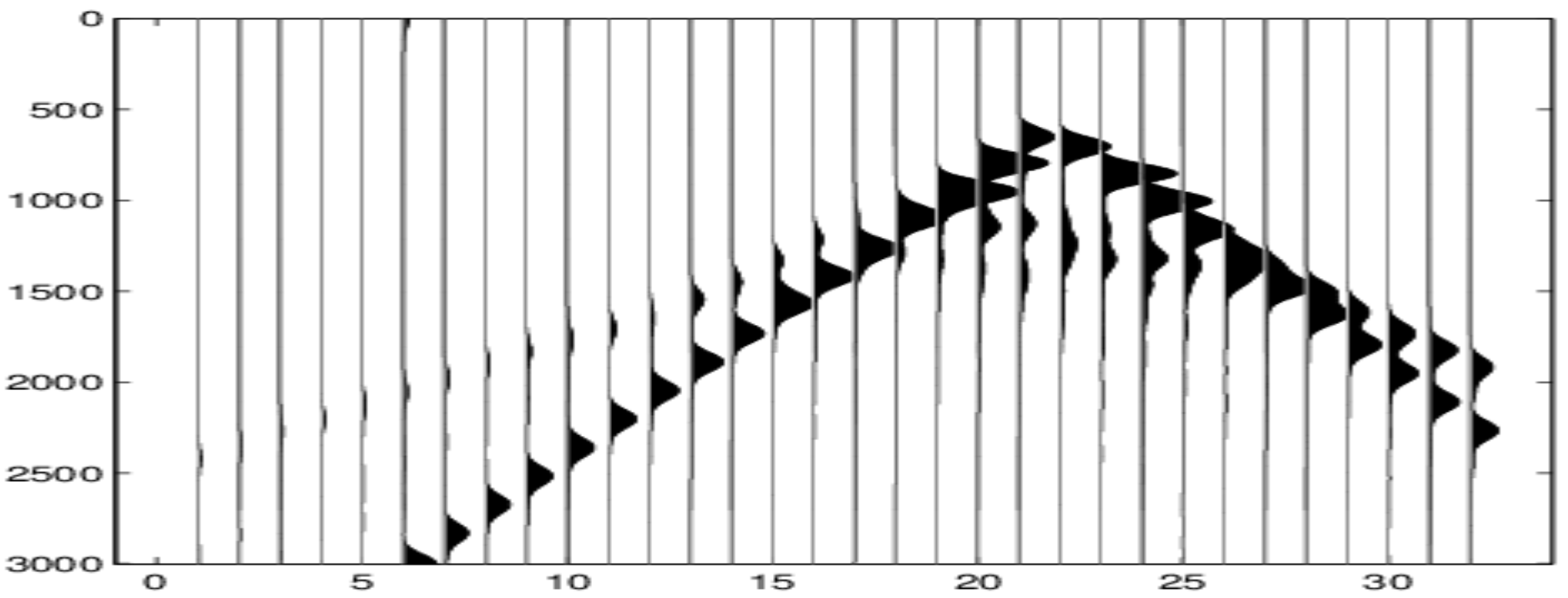

(c)

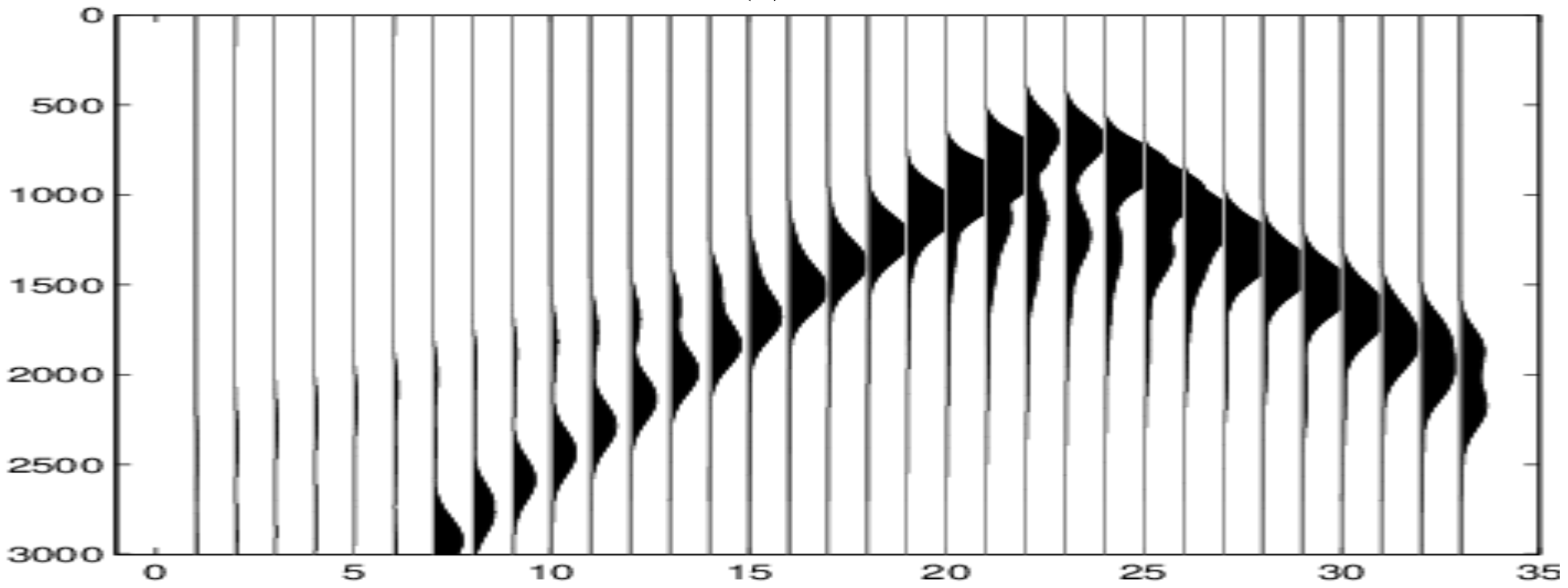

(d)

**Fig.1**: window-averaged envelope (WAE) profiles for the SEG salt model : (a) Original waveform traces; (b) Instantaneous Envelope profile (c) WAE with width 400ms; (d) WAE with width 1000ms.

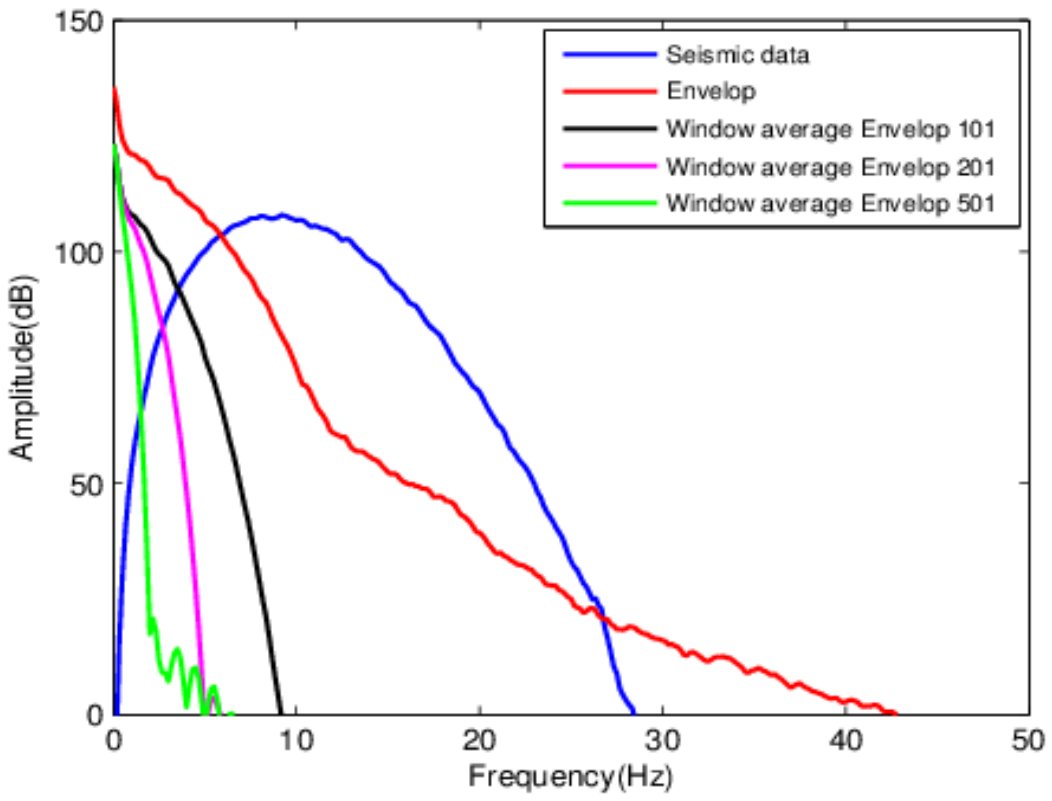

(a)

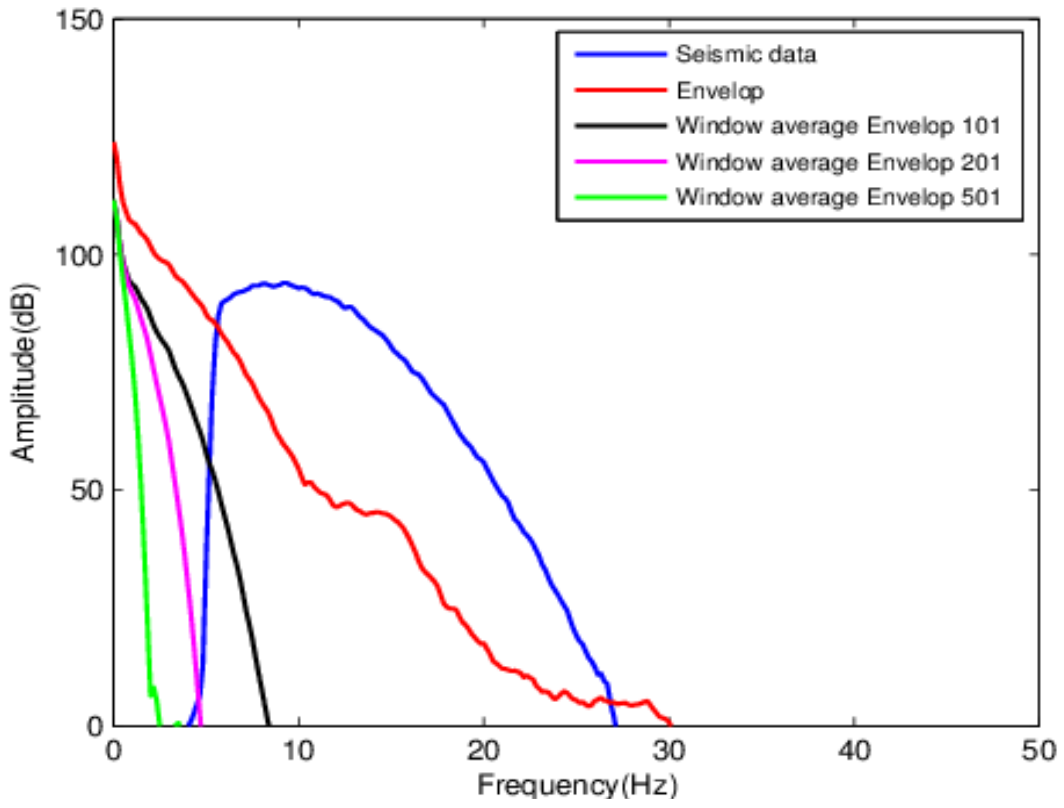

(b)

**Fig.2**: Spectra of WAE with different window widths: red: instantaneous envelope ; black: W=200ms (The window length is 101 and the interval is 2ms); pink: W=400ms (The window length is 201 and the interval is 2ms); green W=1000ms (The window length is 501 and the interval is 2ms). Trace spectra are shown with blue color. (a) For a full-band source; (b) For a low-cut (at 5Hz) source.

Even though the new WAE data have much more ULF information, however, traditional envelope adjoint source cannot take use of this information because of the filtering effects of the adjoint source operator due to the effects of double linearization (see equation (22)). The adjoint source operator will filter and modify the envelope data according to the linearization implied in the chain rule of differentiation, and thus remove much of the useful ULF information. This filtering removes also the deep structural information in the gradient field. Figure 3 shows the comparison of adjoint sources between the conventional FWI (a), traditional envelope inversion (EI) using the waveform Fréchet derivative (b), and the MS-DEI using the new direct envelope Fréchet EFD (c). In Figure 4 we give the corresponding spectra. We know that the adjoint source for conventional FWI is the waveform data residual, so its spectrum is restricted to the effective band of the source spectrum (red spectrum). For traditional EI using the waveform Fréchet Derivative, the adjoint source has a broader spectrum, but the low-frequency components are severely filtered out due to the weak nonlinearity assumption. In contrast, the adjoint source of MS-DEI using the new EFD is just the multi-scale envelope data residual, so the ultra-low-frequency components are preserved in the adjoint source. In addition, from Figure 4, we also see that the adjoint source operator for the traditional EI has also a severe depth filtering effect, since the filter $\left(\partial \mathbf{e}_W / \partial \mathbf{y}\right)^T$ depends on the synthetic seismogram which does not have any deep reflections for the smooth initial model (Figure 4b).

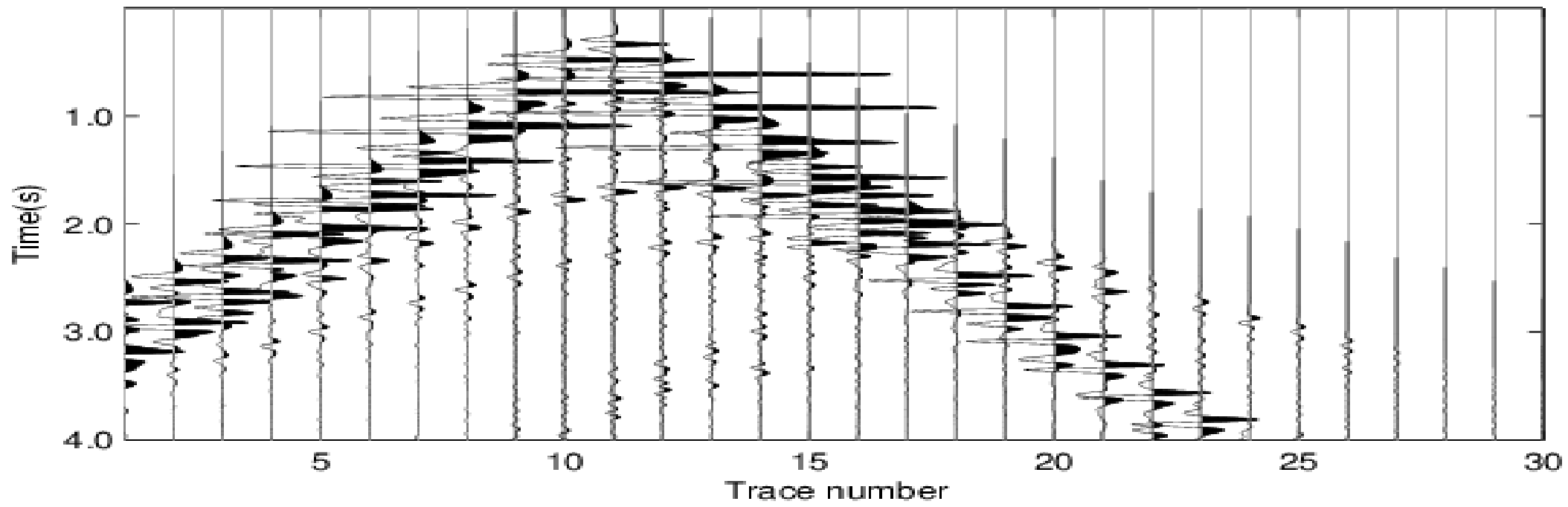

(a)

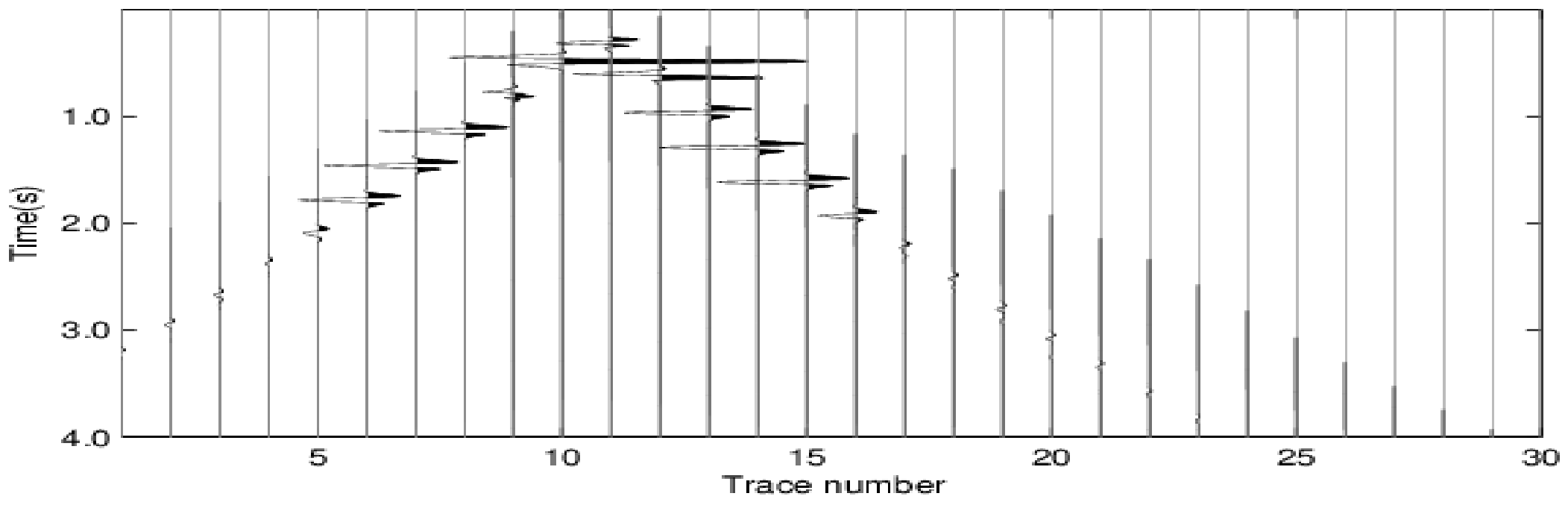

(b)

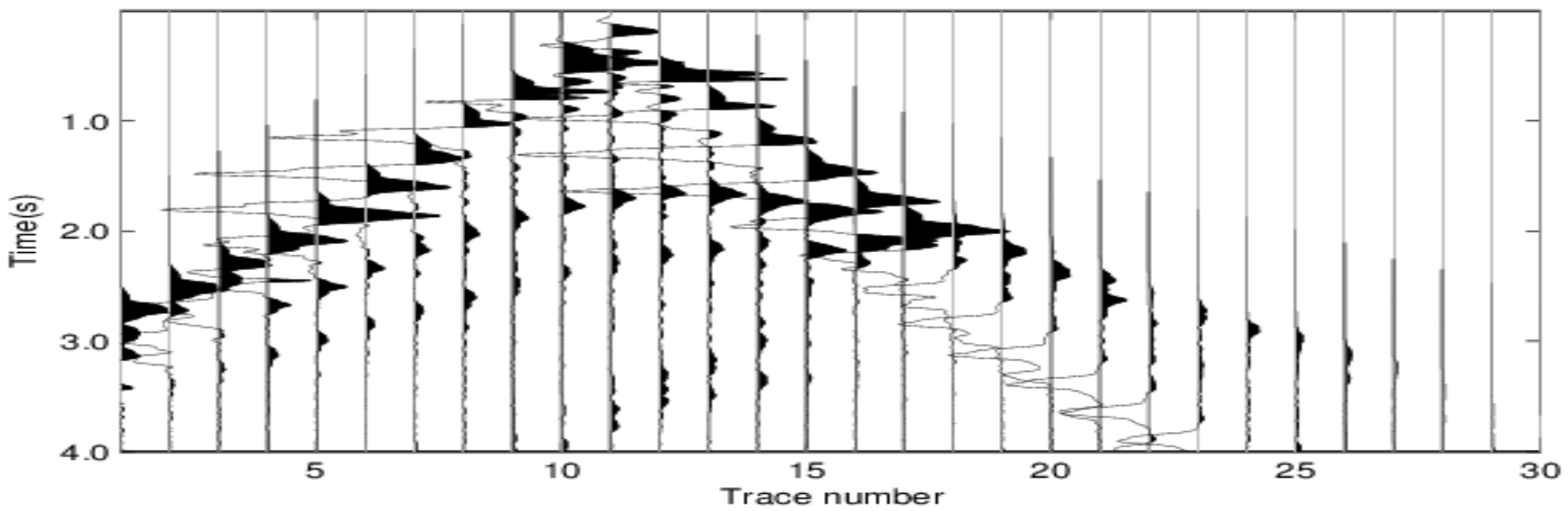

(c)

**Fig.3**: Comparison of adjoint sources between the conventional FWI (a), traditional EI (envelope inversion) using the waveform Fréchet derivative (WFD) (b), and the MS-DEI using the new direct envelope Fréchet derivative (EFD) (c).

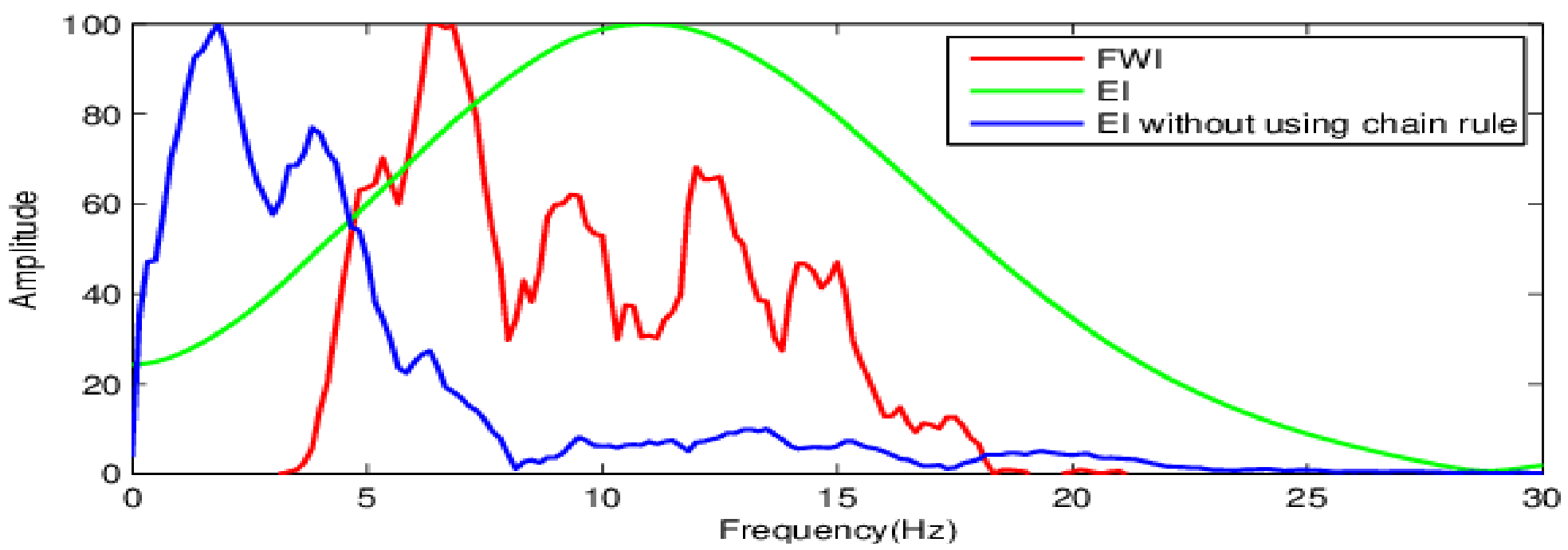

**Fig.4**: Comparison of adjoint sources spectra: FWI (red), conventional EI (green), and MS-DEI using the new envelope Fréchet derivative (EFD) (blue)

## 3 **Joint misfit functional for joint inversion of MS-DEI and FWI**

For multi-scale direct envelope inversion (MS-DEI), its Fréchet derivative is specially constructed for strong scattering (boundary reflections). In this current version, it has some limitations. One shortcoming is the loss of polarity information due to the energy scattering formulation. The other limitation is that its VSO is derived based on the formulation of boundary reflection which is an approximation valid for laterally smooth boundaries. The boundary reflection VSO is good for sharp boundaries, such as the salt boundary, but is not accurate for weak scattering, such the faults or thin-bed reflections. How to combine the weak- and strong-scatterings into a unified formulation is a goal for future research. In our current applications, we mitigate the problem by using a joint inversion with conventional FWI by using a joint misfit functional:

$$\sigma = \frac{1}{2}(1-\lambda)\sum r_u^*(t)\mathrm{r}_u(t) + \frac{1}{2}\lambda\sum \mathrm{r}_e^*(t)\mathrm{r}_e(t). \tag{26}$$

where $\mathrm{r}_u$ is the waveform data residual, $\mathrm{r}_e$ is the envelope (here the window-averaged) data residual, and $\lambda$ is a weighting factor. In the tests performed in this paper, we take $\lambda = 0.5$, which means equal weighting between these two data sets. waveform data have polarity information which can help to improve delineation of the salt bottom and subsalt velocity recovery. Through iterations, FWI can also help correcting the errors made by direct envelope inversion towards the weak scattering objects, such as faults and thin-beds. This is because the window-averaged envelope data will enlarge the thin-layer into thick layers and therefore bring noticeable errors to the background velocity structure. By incorporating FWI inversion into the process, the errors made by MS-DEI can be gradually corrected. This can be seen clearly in the SEG salt model inversion.

## 4 Numerical tests of MS-DEI (multi-scale direct envelope inversion)

In order to test the new envelope Fréchet derivative and its inversion effect on models with strong contrast, we first apply the method to a simple salt-layer model as shown in Figure 5a. To simplify the demonstration, we assume the velocity above the top salt (the surface layer) is known, so we can see the effect of MS-DEI for the salt bottom determination. The initial model is a 1-D linear model with the velocity increasing from 3000m/s to 5000m/s evenly with depth as shown in Figure 5b. The first 250 meters of the initial model are same as the true model. There are 80 shots distributed along the model surface at intervals 100m. For each shot, we use 320 receivers and the interval between receivers is 25m. In order to test the inversion effect of the new method when seismic source lacks low-frequency components, the source is a low-cut wavelet, for which the frequency components below 4Hz are truncated. The dominant frequency of the source is 8 Hz. Figure 6 shows the gradient fields of MS-DEI for different window widths: (a) 600ms width. (b) 300ms width. We can see the scale lengths of gradient field for different window widths. For comparison we also plot the gradient field of conventional FWI in 6(c). Figure 7(a) gives the final inversion result of MS-DEI, and the result of combined MS-DEI plus FWI is shown in 7(b), demonstrating the good recovery of the salt layer velocity and the delineation of the salt bottom. For comparison, we show also the final inversion result of conventional FWI in Figure 7(c). As can be expected, the traditional FWI can only recover the velocity structure near the top salt and the bottom salt is not recoverable.

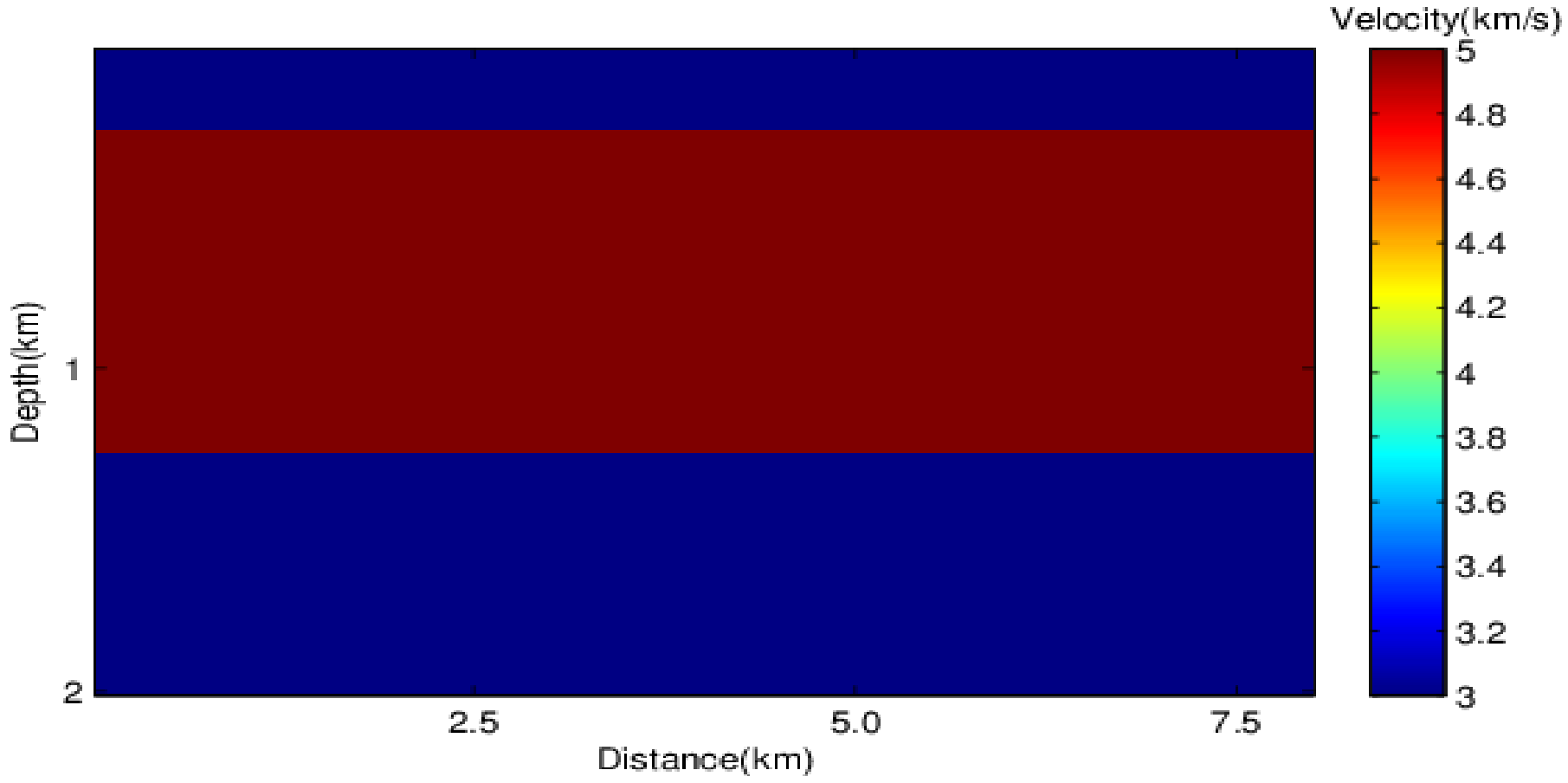

(a)

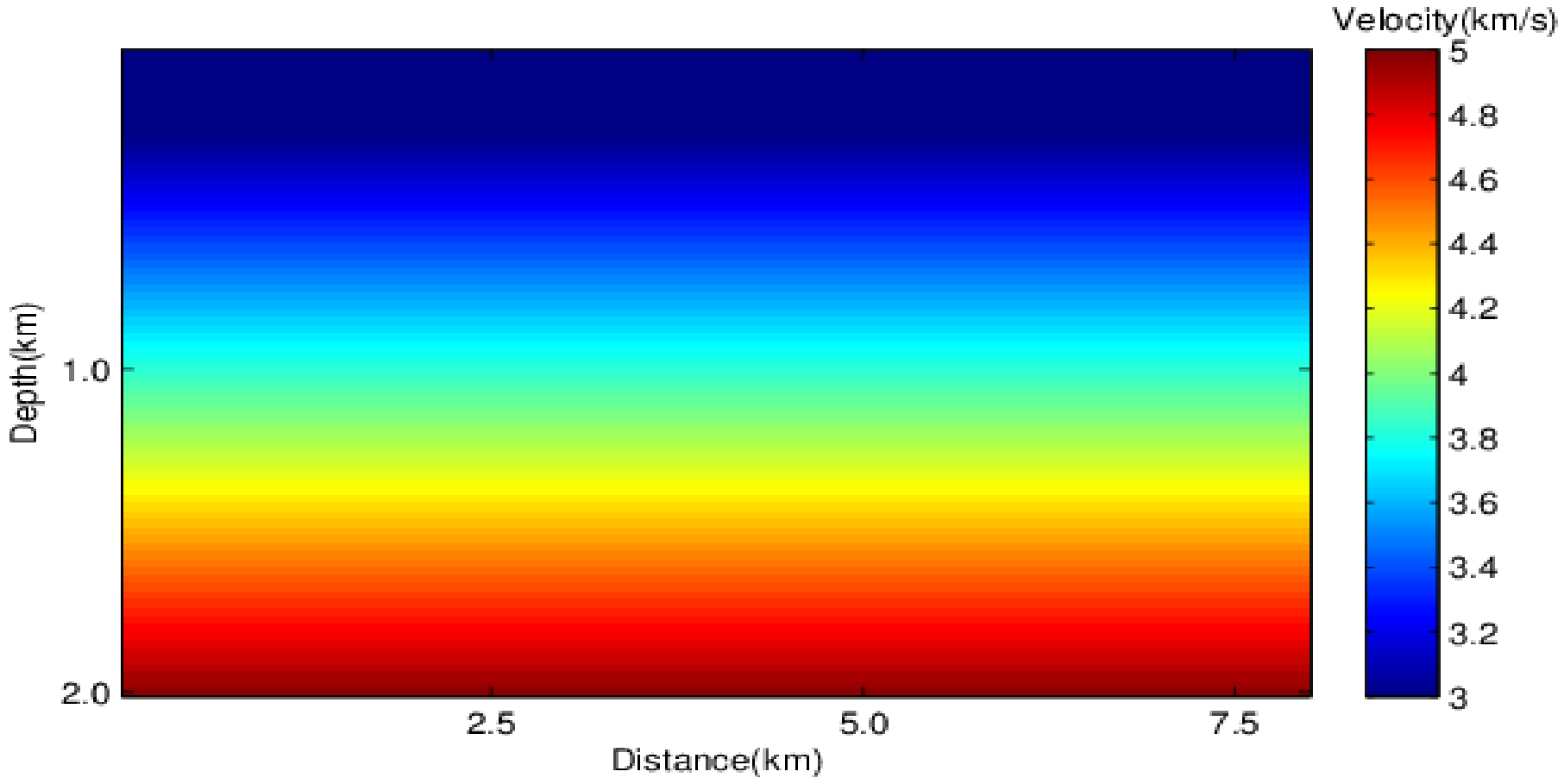

(b)

**Fig.** 5: (a) the test model (a salt-layer model); (b) the initial model for inversion.

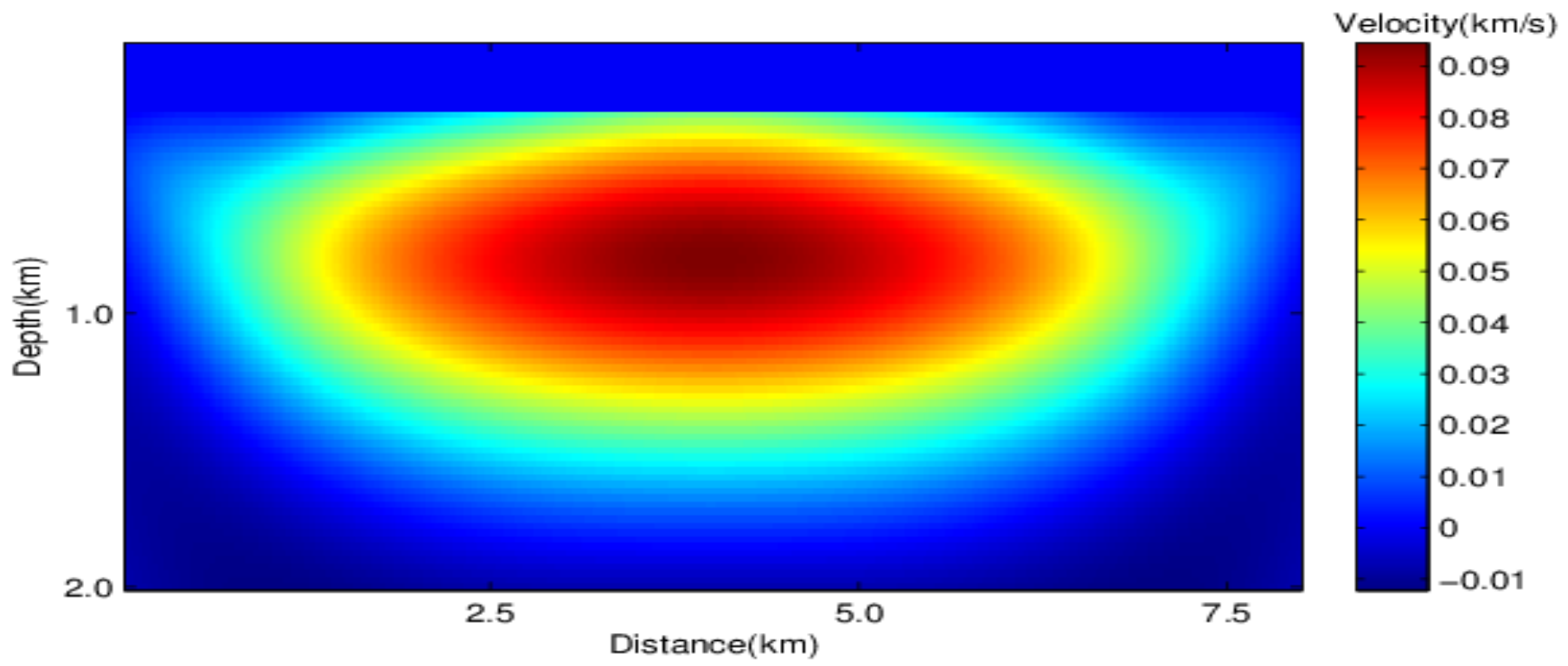

(a)

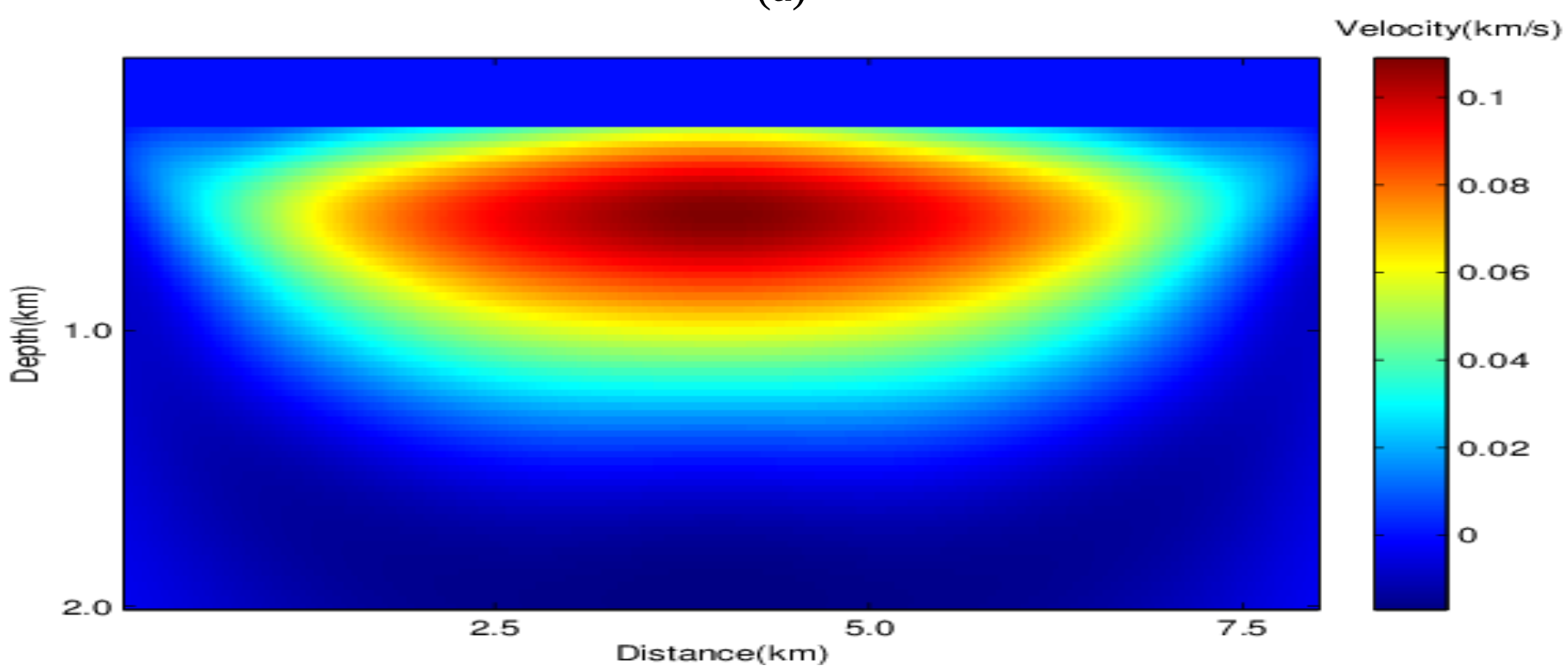

(b)

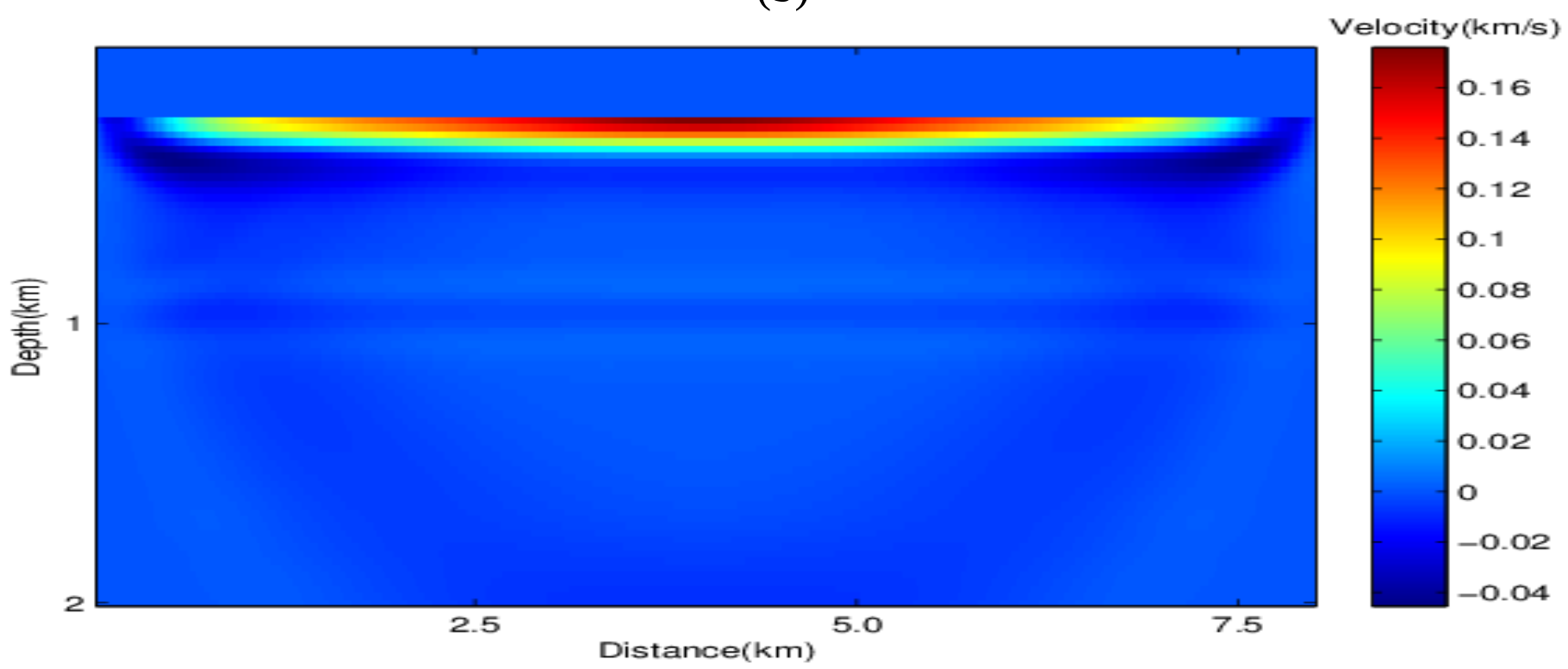

(c)

**Fig.6**: Gradient fields of MS-DEI for different window widths: (a) 600ms width. (b) 300ms width. For comparison we plot the gradient field of conventional FWI in (c).

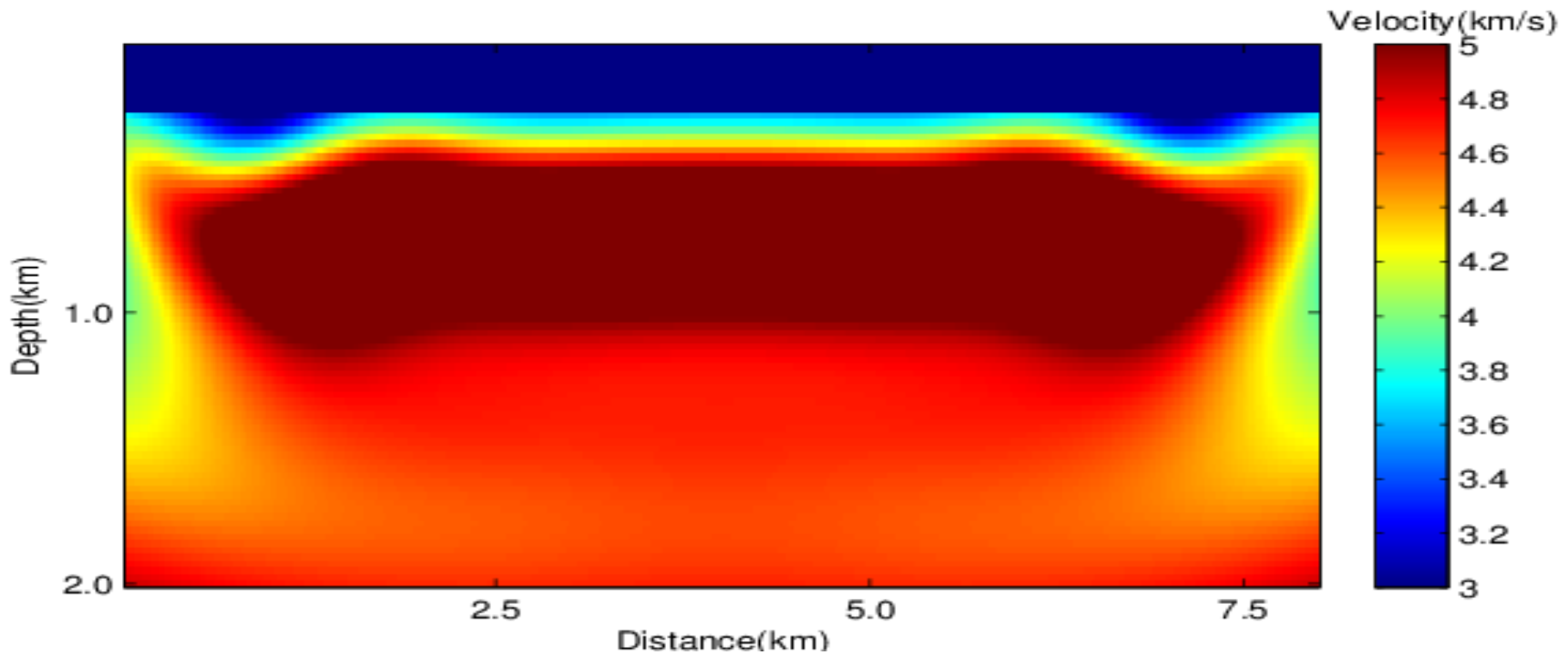

(a)

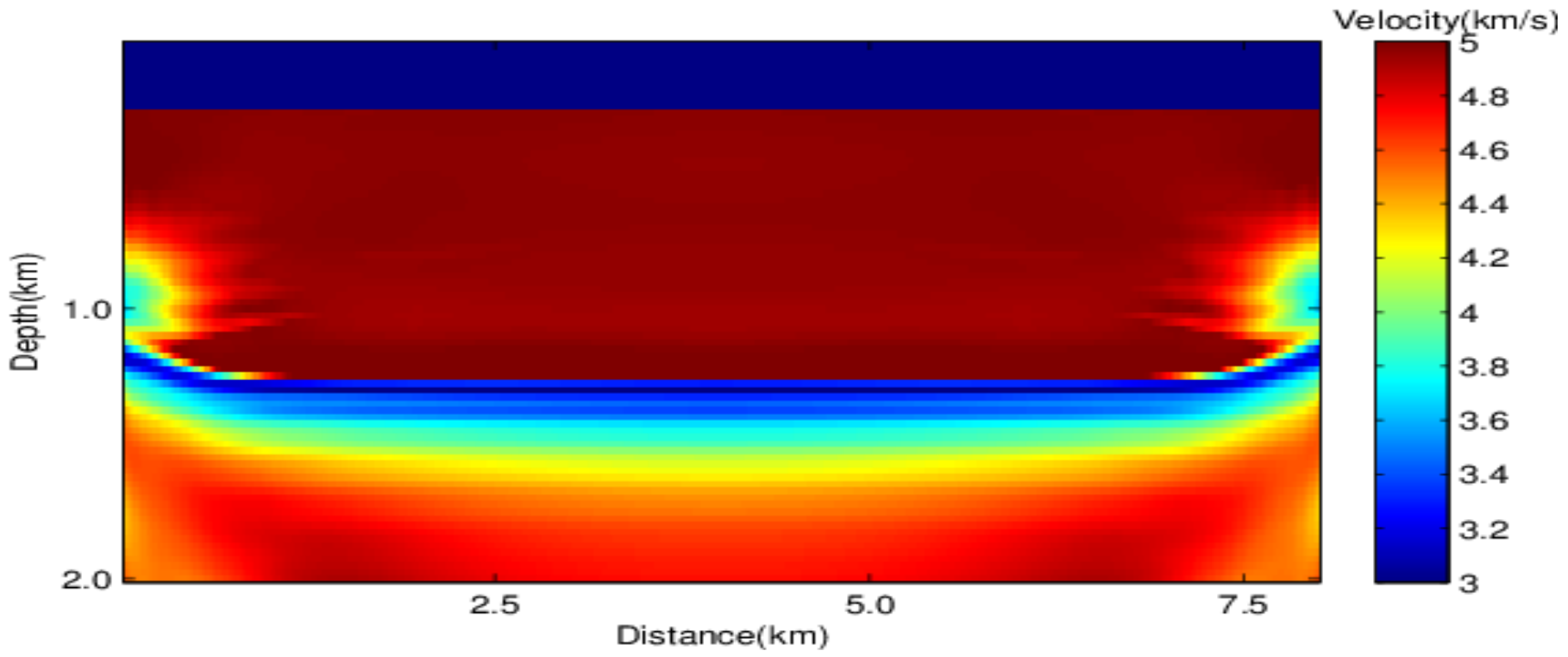

(b)

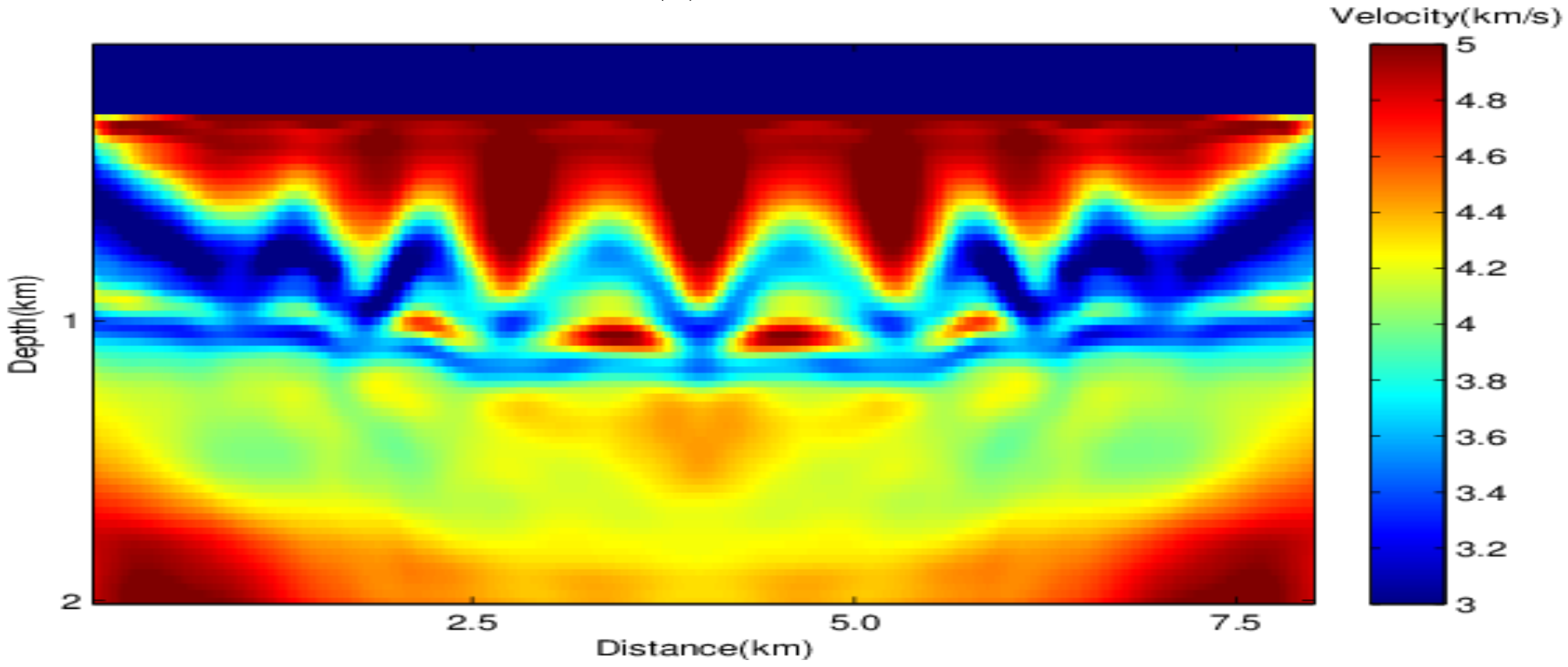

(c)

**Fig.7**: (a) The inversion result of MS-DEI; (b) Inversion result of combined MS-DEI plus FWI; (c) Final inversion result of conventional FWI (for comparison).

For the test using complex model, we show the results on the SEG/EAGE Salt model. The true model is shown in Figure 8a and we set the linear gradient model (Fig. 8b) as the initial model. There are 128 shots distributed along the model surface at intervals of 120m. For each shot, we use 645 receivers with intervals of 24m. A low-cut Ricker wavelet is used as source in the test (cut from 4Hz below, 4-5Hz is the taper zone). The dominant frequency of the source is 9 Hz.

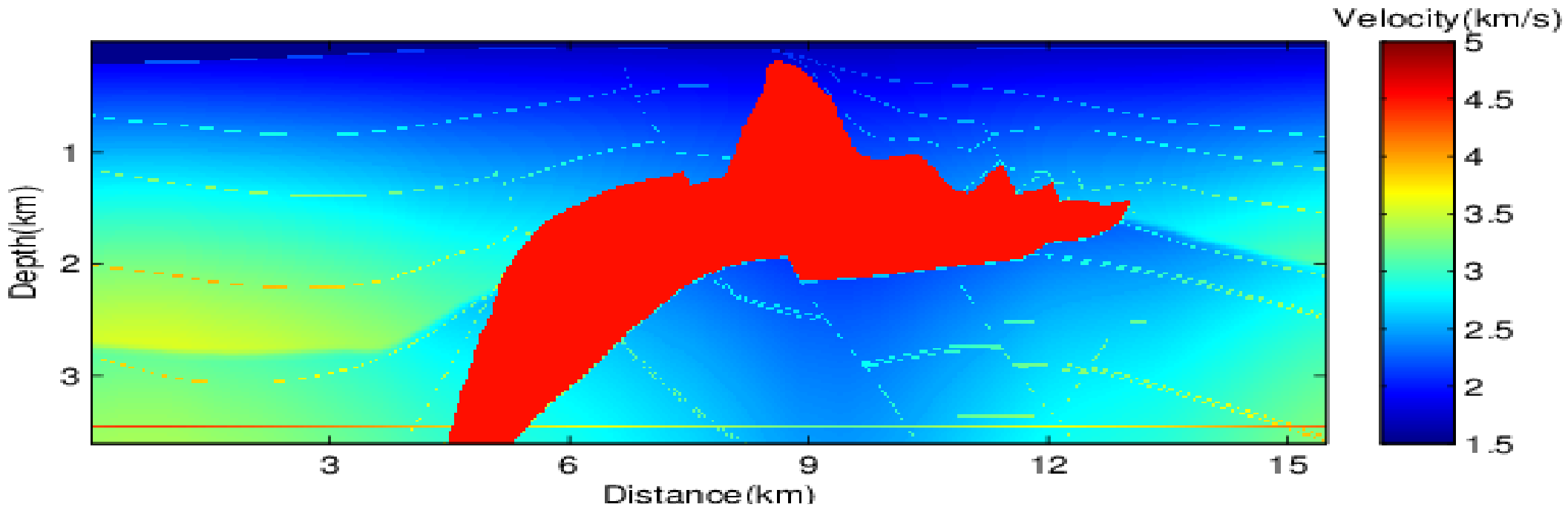

(a)

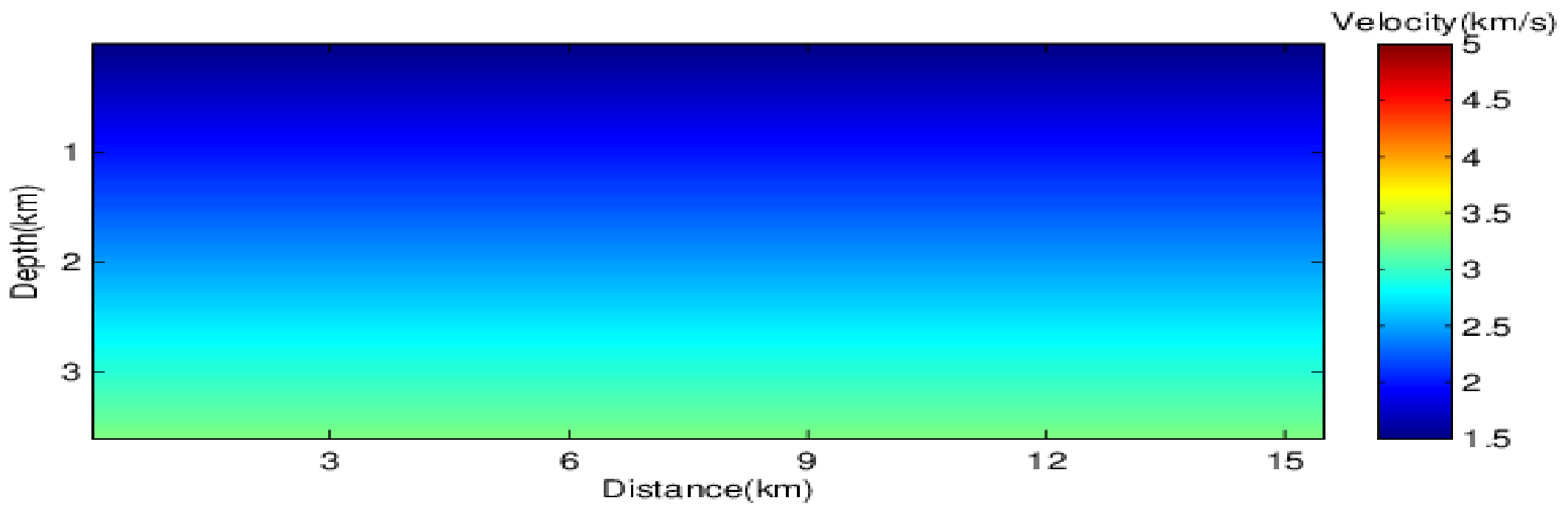

(b)

**Fig.8**: SEG/EAGE salt model (a) and the initial model (b) for inversion.

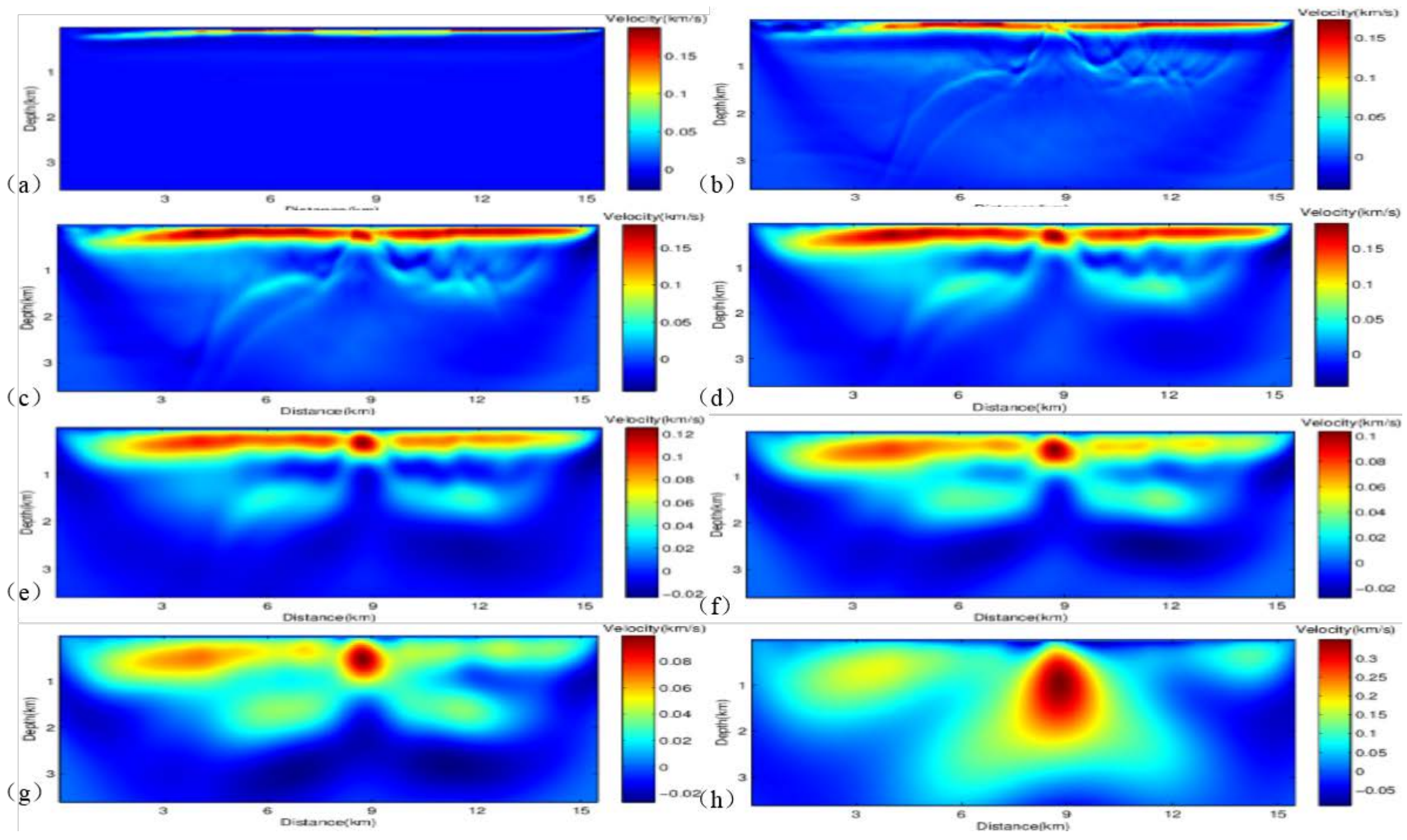

**Fig.9**: Gradient fields using different Fréchet derivatives: (a) conventional EI using waveform Fréchet (WFD); (b)-(h) MS-DEI using EFD (envelope Fréchet derivative) with different window-widths: (b) original, (c) 20ms, (d) 50ms, (e) 100ms (f) 200ms, (g) 300ms, (h) 500ms.

Next, we show the difference in gradient fields between using the new EFD and using the traditional WFD for envelope data. Figure 9 shows the Gradient fields using different Fréchet derivatives: (a) conventional EI using waveform Fréchet; (b)-(h) MS-DEI using envelope Fréchet with different window-widths: (b) original, (c) to (h) are 20ms, 50ms, 100ms, 200ms, 300ms, and 500ms, respectively. We see that for this strong-nonlinear case of salt model inversion, using the traditional waveform Fréchet derived by the chain rule results in a gradient similar to the case of FWI, and the gradient reaches only shallow depth; while the new envelope Fréchet can reach greater depth, since the adjoint source in the latter case is just the envelope residual. The gradient field is the backpropagated envelope residual multiplying by the virtual source term. We know that the gradient field is a modified migration image field (here is modified by a factor of virtual source term) (e.g. Lailly, 1983; Mora 1989; Pratt *et al*., 1998). From wave physics, equation (24) involves energy backpropagation and imaging. Even though energy packet focusing is not efficient due to the lack of interference, but the traveltime stacking can give us the dull image we needed for large-scale recovery. In contrast, we see from equation (5) and Figure 3 that effective data residual, i.e. the adjoint source, using the traditional Fréchet (WDF) is severely distorted by a filter $\partial e / \partial u$ which has a devastating effect on envelope residual. For the multi-scale envelope data, the new Fréchet (EFD) depicts the better linear correspondence between the multi-scale data and the multiple-scale velocity structures (Figure 9 b-h). Figure 10 gives the multiple-scale gradient field of MS-DEI by superposing the individual gradient fields, showing how the gradient field resembles the gross feature of the salt structure. Compared with the gradient field of EI using the traditional Fréchet (WFD) (Figure 9a), the superiority of MS envelope Fréchet is obvious.

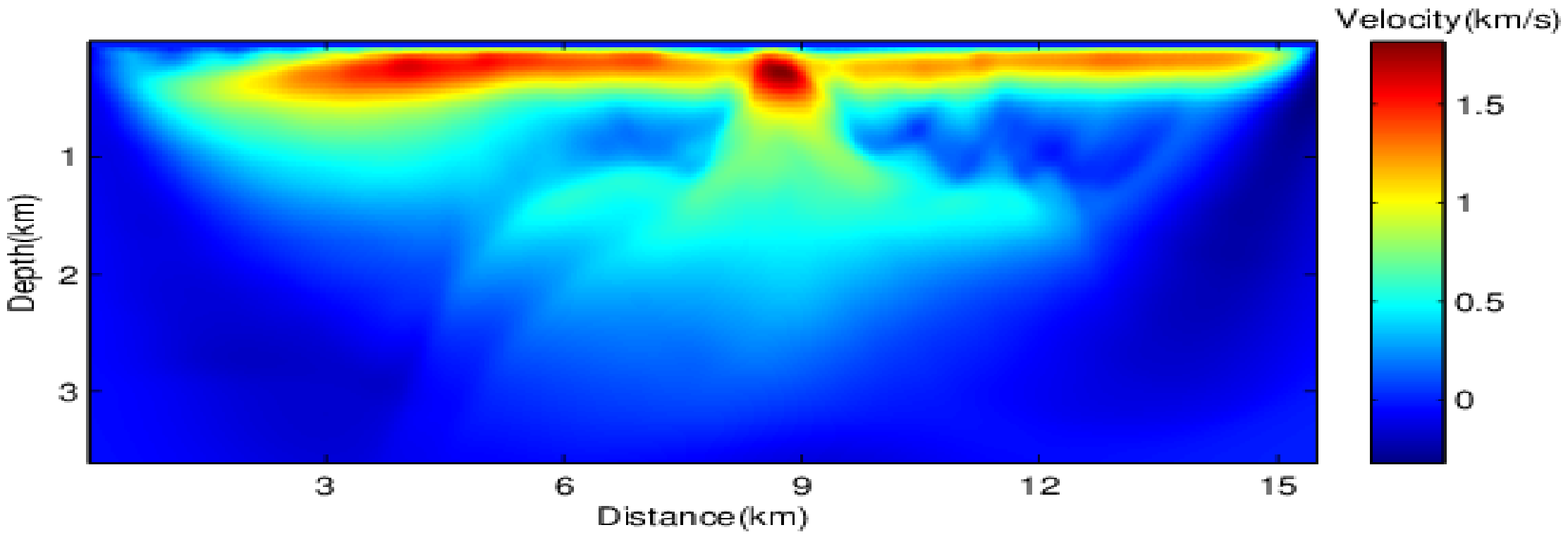

**Fig.10**: Gradient field of MS-DEI with multi-scale envelope data (window widths 0-500ms) (see figure 9). Note that the waveform data are produced with a low-cut (from 4Hz below) source wavelet.

Now we show the inversion results of our MS-DEI using the new direct envelope Fréchet applied on the SEG 2D salt model. For comparison, we plot the results of conventional FWI in Fig. 11(a) and conventional EI+FWI in Fig. 11(b). Due to the strong-contrast and large size of the salt body, the traditional FWI can only see the top salt boundary. For the same reason, conventional envelope inversion cannot penetrate deep into the salt body. The conventional EI still has cycle-skipping problem although less severe than FWI. Secondly, conventional EI derived the gradient using the chain rule of differentiation and updates the model using the waveform Fréchet derivative which is based on weak-scattering assumption. In this way, many severe limitations of the first order waveform Fréchet is expected to bring difficulties to the conventional EI or EI+FWI. In sharp contrast, we plot the results of MS-DEI with different window widths in Fig.12. We see that large scale structure has been recovered due to the use of the new envelope Fréchet and the multi-scale decomposition of envelope data.

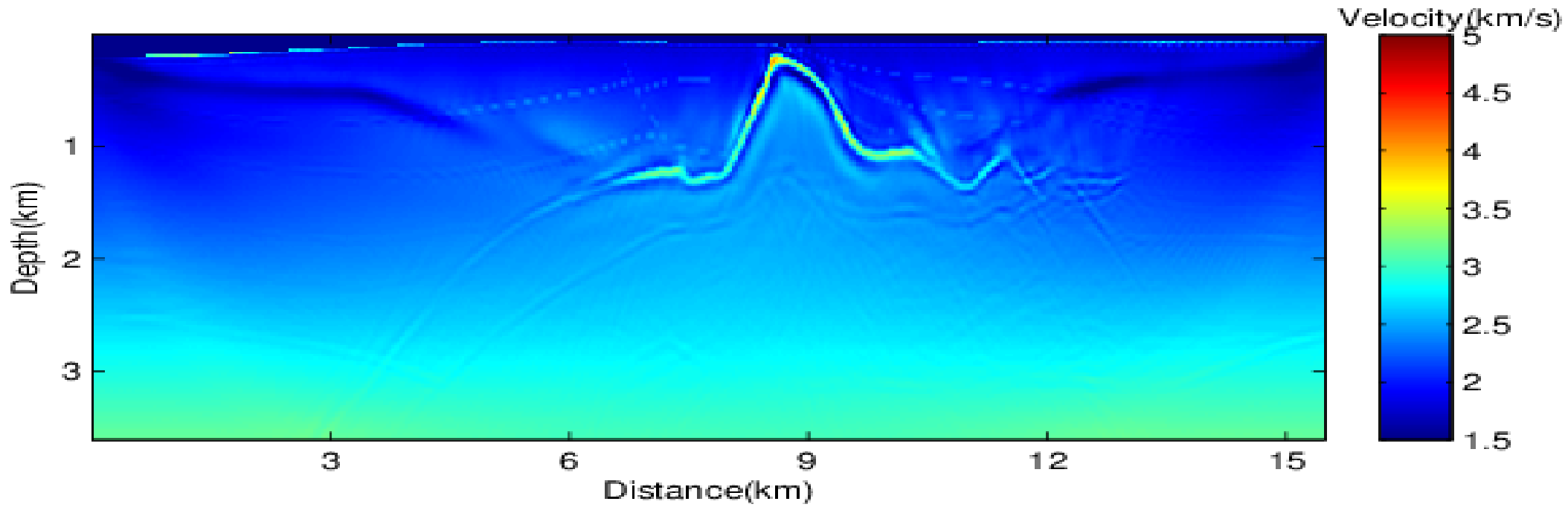

(a)

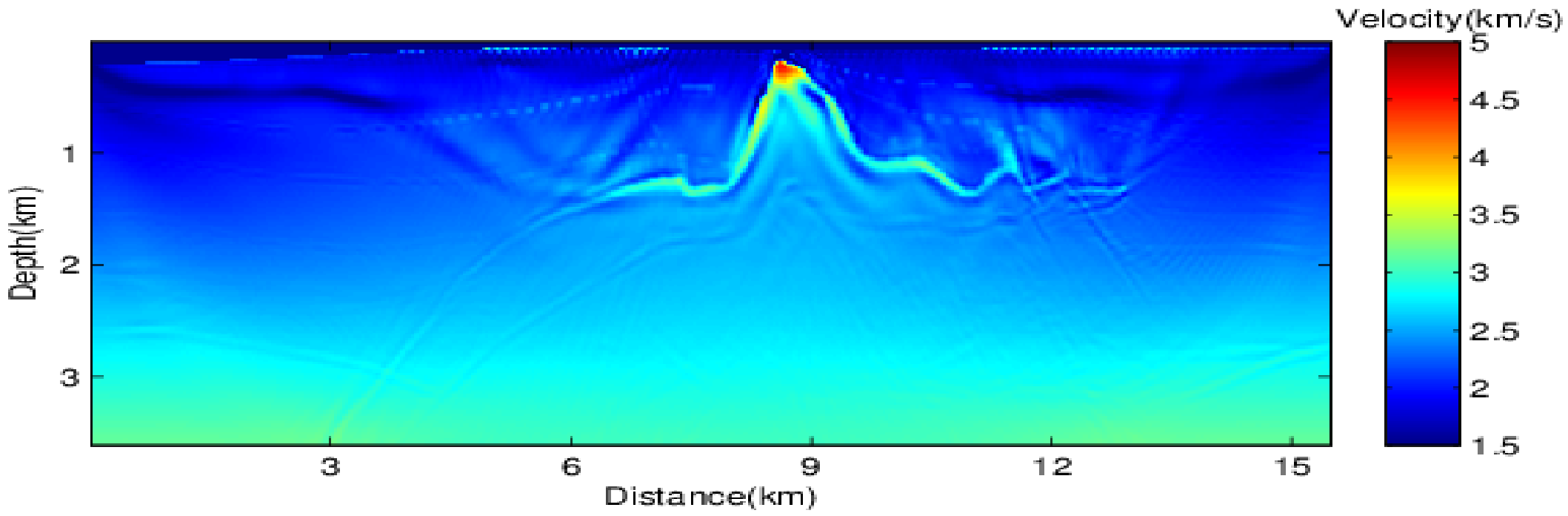

(b)

**Fig.11**: Inversion results of (a) conventional FWI; (b) Conventional EI+FWI

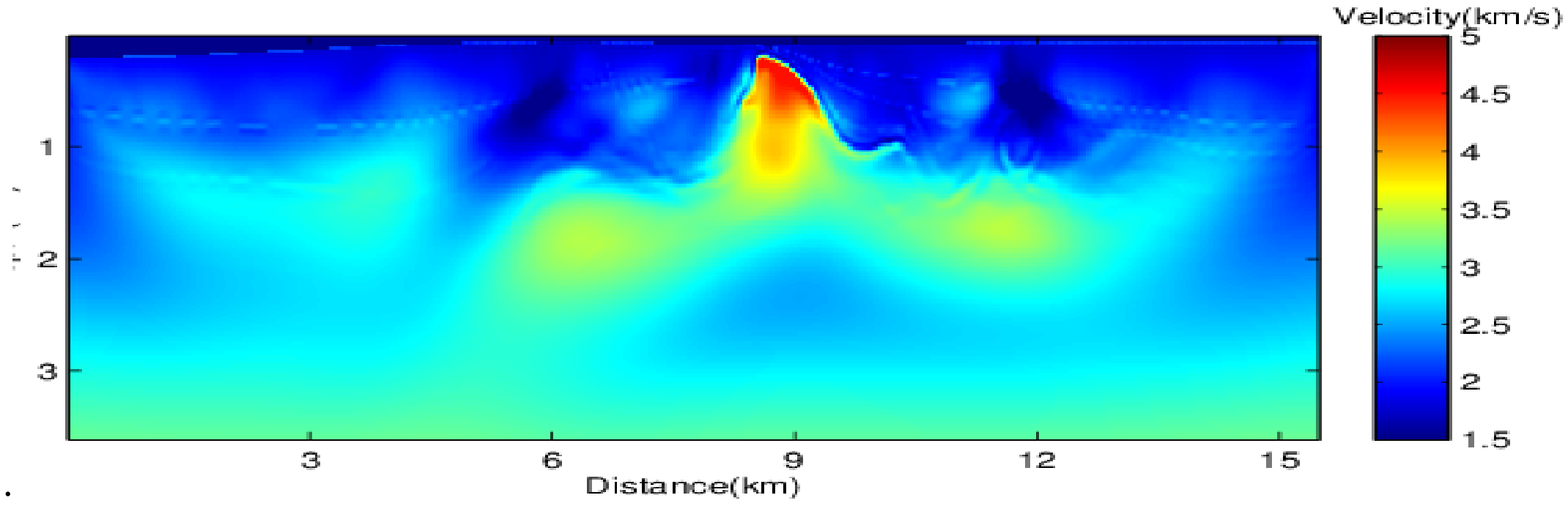

(a)

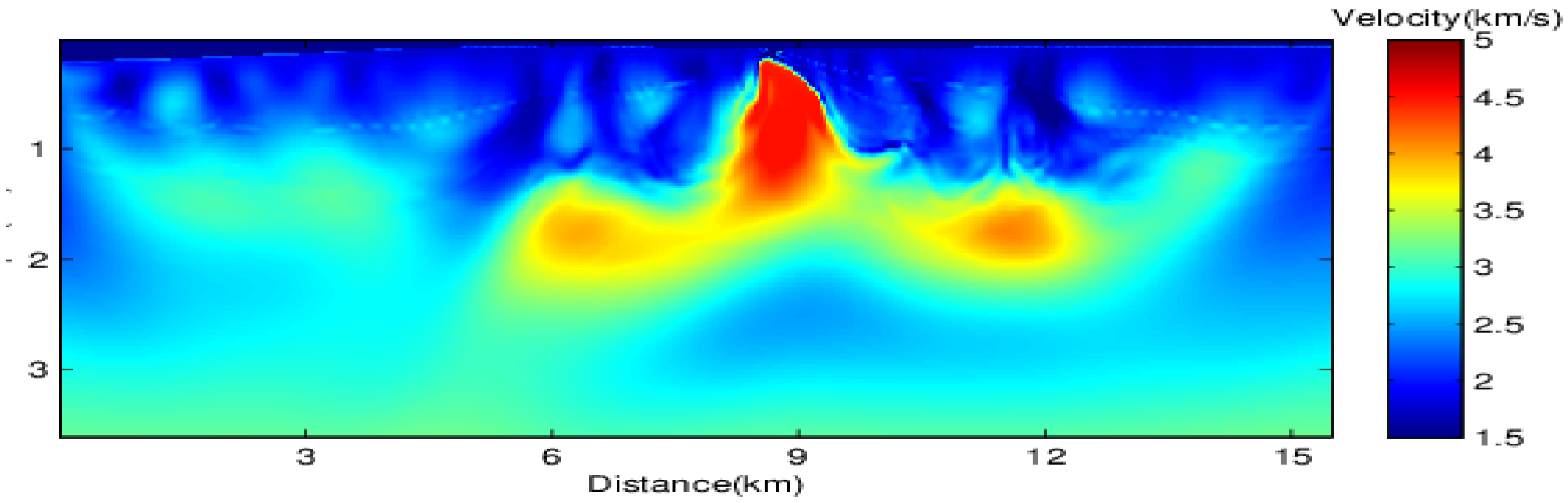

(b)

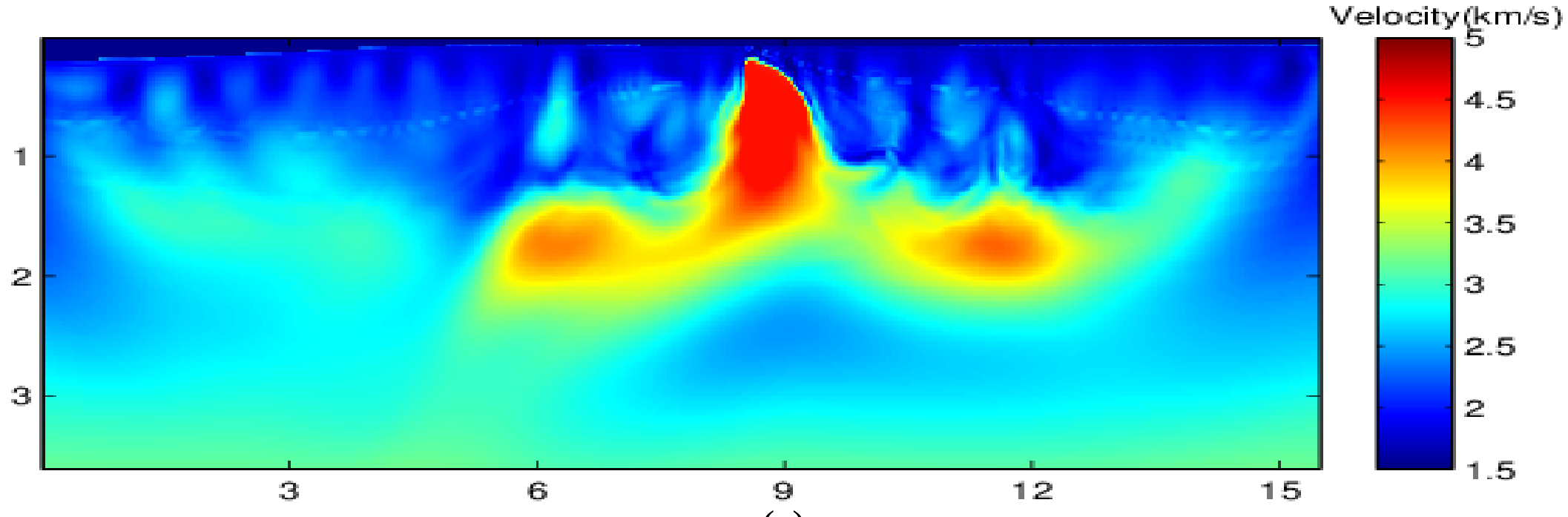

(c)

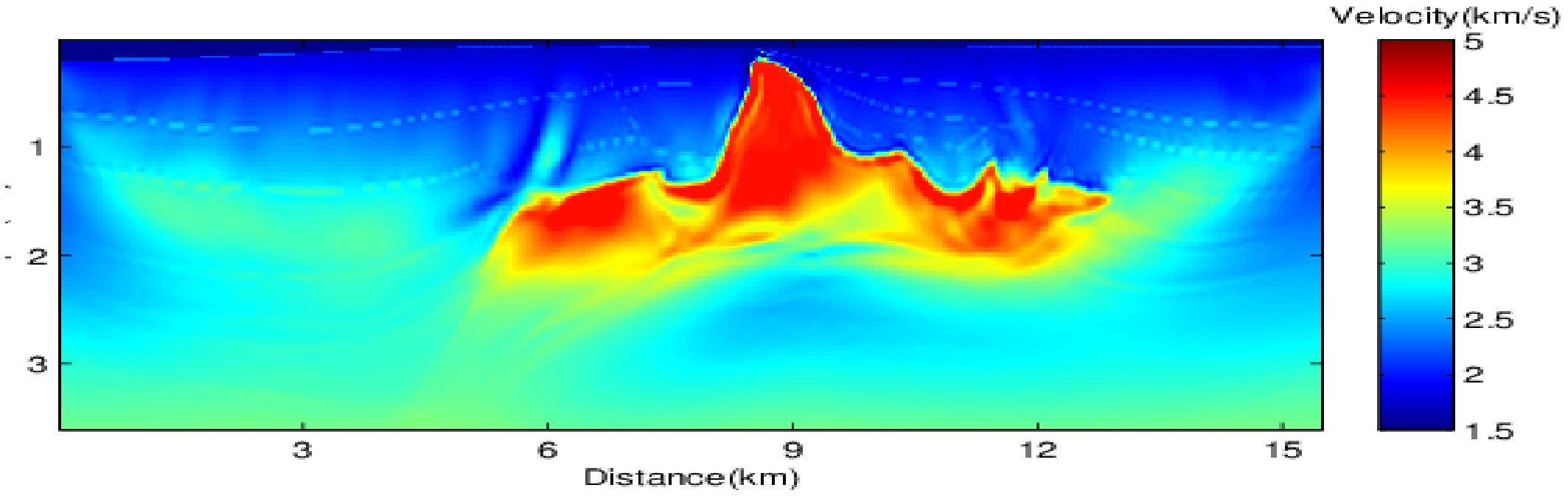

(d)

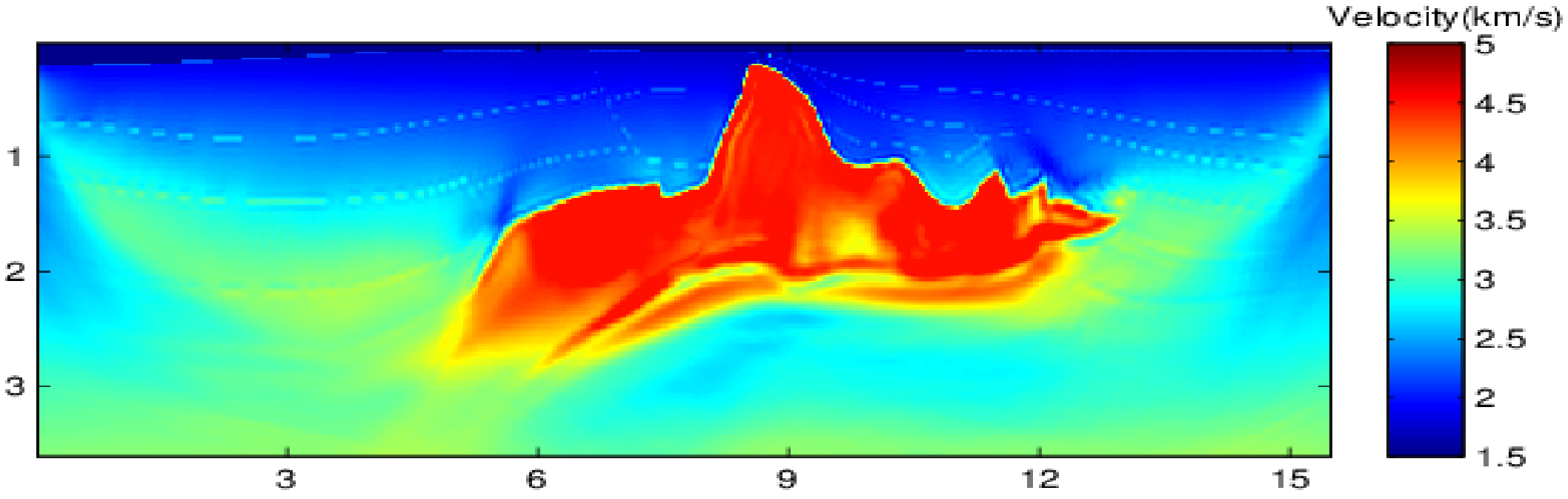

(e)

**Fig.12**: MS-DEI (multi-scale envelope inversion) results at different stages of inversion. The source is a low-cut (cut from 4Hz below) Ricker wavelet: (a) using window-width $W = 300$ ms, (b) $W = 150$ ms, (c) $W = 50$ ms. For each scale the results of previous scale (larger scale) is used as initial model. (d) is the result after applying FWI to the result in (c). (e) final result after three loops of iterations (each loop consists of iterations with different window lengths).

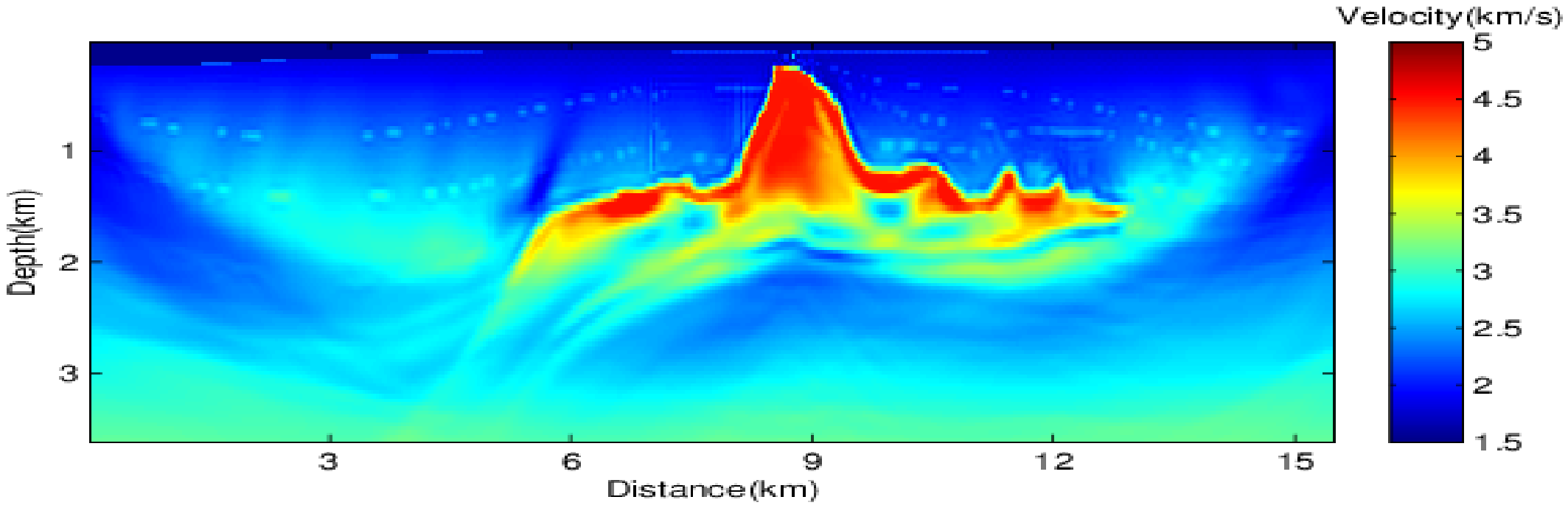

(a)

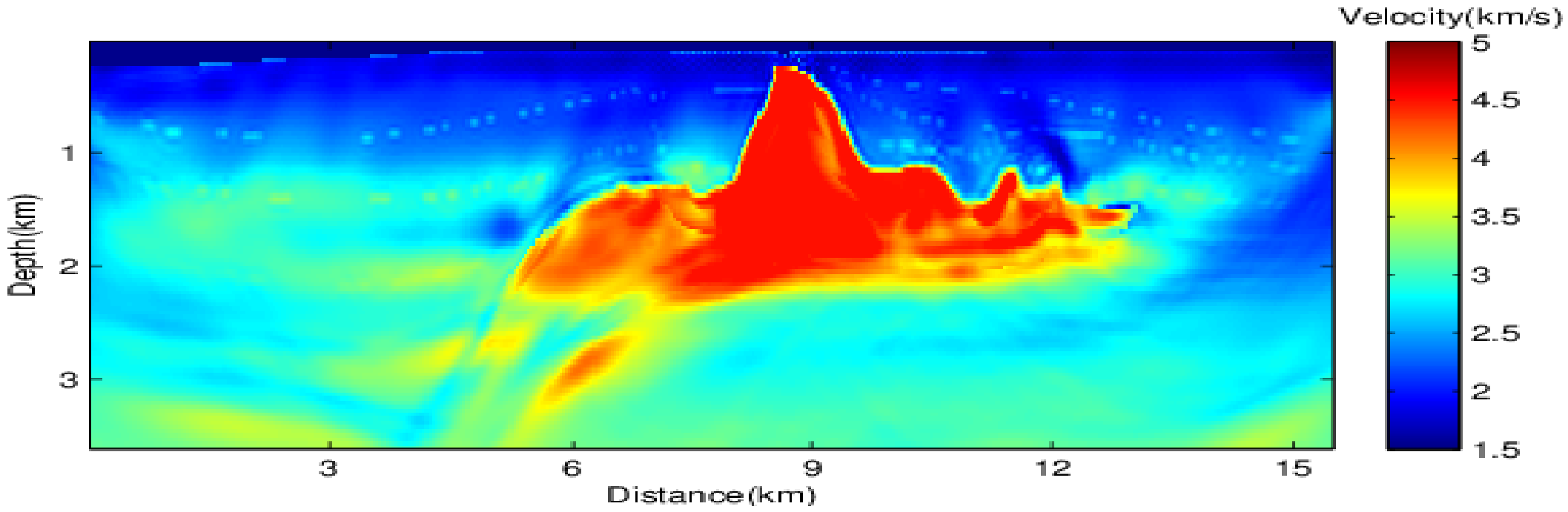

(b)

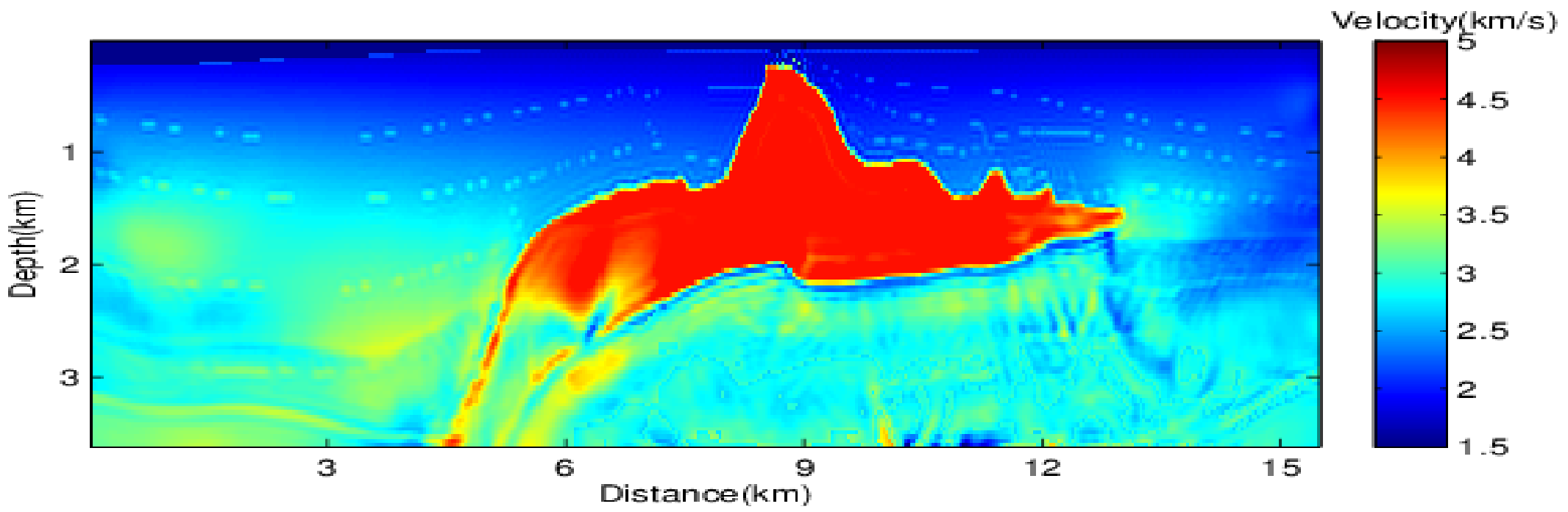

(c)

**Fig.13**: MSDEI + FWI combined with multi-offset method with three loops of iterations: (a) first loop; (b) second loop; (c) third loop. (each loop consists of iterations with different window lengths).

In order to recover the detailed structure of the salt top, we apply the regular FWI to each window of the MS-DEI. Fig. 12(d) shows the result after FWI applied to the first loop of multi-window iteration. The same procedure is applied to the second and third loops and the final result of the third loop is shown in Fig. 12(e). We see that the salt body is gradually recovered. However, the salt bottom is still contaminated by artifacts. To improve the salt bottom delineation, we incorporated the multi-offset method into our MS-DEI. First we use the long-offset envelope data (mainly transmitted waves) to recover the low-wavenumber structure. This is helpful for the recovery of subsalt background velocity if the salt flank has been delineated; and then in later loops we turn to the mid- and short-offset envelope data to refine the salt bottom using reflected waves (for details see Chen et al., 2017). The final results are shown in Fig. 13. We see that not only the salt bottom is better recovered, but also the smooth background structure is improved. Nevertheless, the deep part of the salt flank has not totally recovered and the subsalt structure is also not accurate. The is due to the poor illumination of deep steep flank and weak strength of reflection signals beneath the salt dome. We will continue the theory and method development and combine other techniques of subsalt imaging and inversion to further improve the results.

## 5 Discussion and Conclusion

Multi-scale direct envelope inversion (MS-DEI) has two key ingredients: one is the multi-scale (MS) decomposition of envelope curves, the other is the introduction of a new envelope Fréchet derivative (EFD) operator and a gradient method based on the operator. The MS decomposition of envelope data is very different from the decomposition of waveform data. Envelope extraction is a nonlinear signal processing which changes the waveform data (received scattered wavefield) into envelope data, which records the scattered energy packets. These energy packets do not carry phase information so no interference will happen when you decompose or superpose these packets. This is why we can perform linear filtering to get the low-frequency components of these packets. On the contrary, linear filtering of waveform data can only obtain the low-frequency components limited by the source spectra. The other key ingredient is the new direct envelope Fréchet derivative (EFD). Since traditional way to get the Fréchet derivative for a new data functional is the use of chain rule of differentiation which implies the weak perturbation and the linearization of both the waveform Fréchet derivative and the envelope functional derivative. In the case of strong-nonlinear inversion, such as in the salt structure inversion using a linear 1-D initial model, the nonlinearity of both functional derivatives are so strong such that the linearization becomes invalid. On the other hand, the direct envelope Fréchet (EFD) has a much better linearity with respect to the velocity variations. Therefore, the new EFD plays a critical role in MS-DEI. We demonstrated both analytically and numerically that the new theory/method of MS-DEI can be applied to the case of strong-nonlinear inversion for large-scale strong-contrast media.

**Acknowledgments**.

We thank Morten Jakobsen, Yingcai Zheng, Tariq Akahalfa, Thorne Lay, Jingrui Luo, Pan Zhang, Yuqing Wang and Yong Hu for helpful discussions. This work is supported by WTOPI (Wavelet Transform On Propagation and Imaging for seismic exploration) Research Consortium and other funding resources at the Modeling and Imaging Laboratory, University of California, Santa Cruz. We are grateful to our Consortium sponsors for their financial support and allowing us to publish our research results.

## Appendix A: Nonlinear sensitivity operator of waveform data to velocity perturbation

In full waveform inversion, Fréchet derivative plays a key role, which can be viewed as a linearized sensitivity operator of waveform data to medium velocity (or other parameters) perturbation. We know that on the real Earth, the wave equation is strongly nonlinear with respect to the medium parameter changes, except for short range propagation in weakly heterogeneous media. It is known that Fréchet derivative corresponds to a sensitivity operator with the Born approximation (Tarantola, 2005) and its application is severely limited by this approximation. Wu and Zheng (2013, 2014) proved that higher order Fréchet derivatives can not be neglected in the strong-nonlinear case and introduced the *nonlinear sensitivity operator* (NLSO) which corresponds to the *nonlinear partial derivative* operator (NLPD) for the acoustic wave equation. In this paper we consider only the velocity inversion and NLPD becomes a *nonlinear functional derivative* (NLFD). In this appendix, we summarize briefly the concept and derivation of NLSO to facilitate the understanding of the problem in applying the chain rule of differentiation involving nonlinear sensitivity operators.

A waveform data set can be defined as a function $\mathbf{u}(x_g, x_s, t)$ in the data space $\mathbf{D}$, which can be related to the model function $\mathbf{m}(\mathbf{x})$ in the model space $\mathbf{M}$ by an operator (mapping)

$$\mathbf{u} = \mathbf{A}(\mathbf{m}) \tag{A1}$$

Assume an initial model $\mathbf{m}_0$, we want to quantify the sensitivity of the data change $\delta\mathbf{u}$ to the model change $\delta\mathbf{m}$. The total nonlinear sensitivity operator can be expressed by a Taylor series including all orders of derivative (Wu and Zheng, 2014, equation 27)

$$\begin{aligned}\delta\mathbf{u}(\delta\mathbf{m},\mathbf{m}_0) &= \mathbf{u}(\mathbf{m}_0+\delta\mathbf{m}) - \mathbf{u}(\mathbf{m}_0) = \mathbf{A}(\mathbf{m}_0+\delta\mathbf{m}) - \mathbf{A}(\mathbf{m}_0)\\ &= \mathbf{A}'(\mathbf{m}_0)\delta\mathbf{m} + \frac{1}{2!}\mathbf{A}''(\mathbf{m}_0)(\delta\mathbf{m})^2 + \cdots + \frac{1}{n!}\mathbf{A}^{(n)}(\mathbf{m}_0)(\delta\mathbf{m})^n + \cdots\\ &= \mathbf{A}^{(NL)}\delta\mathbf{m}\end{aligned} \tag{A2}$$

where $\mathbf{A}'$, $\mathbf{A}''$, and $\mathbf{A}^{(n)}$ are the first, second, and nth order Fréchet derivatives and $\mathbf{A}^{(NL)}$ is the nonlinear functional derivative defined as

$$\mathbf{A}^{(NL)} = \mathbf{A}'(\mathbf{m}_0) + \frac{1}{2!}\mathbf{A}''(\mathbf{m}_0)(\delta\mathbf{m}) + \cdots + \frac{1}{n!}\mathbf{A}^{(n)}(\mathbf{m}_0)(\delta\mathbf{m})^{n-1} + \cdots \tag{A3}$$

From the above equation, we see that ***the nonlinear sensitivity operator is model-dependent and perturbation-dependent***. Stronger the perturbation is, more important the higher order terms are. Series (A2) is closely related to the Born series. In traditional seismic inversion, normally higher order Fréchet derivatives are neglected. However, the neglect of higher order terms leads to the Born approximation! In order to go beyond the Born approximation, we need a sensitivity operator including all the higher order Fréchet derivatives.

In this paper, we will adopt the notations of functional and functional derivative used in physics literature (e.g Engel and Dreizler, 2011, See also “Functional derivative” in Wikipedia), since it expresses more explicitly the functional dependence. In notation of functional, (A1) can be rewritten as

$$u(x_g, x_s, t) = A[(m(\mathbf{x})](x_g, x_s, t) \tag{A4}$$

where $\mathbf{x} = (x, y, z)$ and $x_g$ , $x_s$ and $t$ are the receiver, source locations and time, respectively; $A[.](.)$ is a functional and, [.] and (.) specify the *functional dependence* and the *function dependence*, respectively.

Now we study the waveform data change (variation) $\delta u$ as a functional dependence of the medium (such as the velocity field) variation $\delta m$, which is usually quite nonlinear as in (A3):

$$\left(\frac{\delta u}{\delta m}\right)_{NL} = A^{(NL)} = \frac{\delta u}{\delta m} + \frac{1}{2!}\frac{\delta^2 u}{\delta m^2}\delta m + \cdots + \frac{1}{n!}\frac{\delta^n u}{\delta m^n}(\delta m)^{n-1} + \cdots\cdots \tag{A5}$$

(for mathematical definitions of higher order Fréchet derivatives and Taylor expansion of nonlinear functional, see Teschl, 1998; Zhang, 2005). We call $(\delta u / \delta m)_{NL}$ the *nonlinear sensitivity functional* or *nonlinear functional derivative*. Neglecting the higher order terms leads to the linearization of the nonlinear functional $(\delta u / \delta m)_{NL}$ so that

$$\left(\frac{\delta u}{\delta m}\right)_{NL} \approx A^{(Linear)} = \frac{\delta u}{\delta m} \tag{A6}$$

However, this linearization is only valid in the case of *weak nonlinearity*. We see that $(\delta u / \delta m)_{NL}$ is perturbation-strength $|\delta m|$ dependent, and only its linear approximation $\delta u / \delta m$ is perturbation-strength independent. In case of strong perturbation, the linear approximation may become invalid.

## Appendix B: Chain rule of sensitivity operators and linearization of functional derivatives

To demonstrate the difficulty and problem of using linearized functional derivative (sensitivity operator) to handle strong-nonlinear functional, we further analyze the chain rule of differentiation in the case of strong nonlinear sensitivity.

As we show in (A4), the waveform data $u(\mathbf{a})$ is a function in data space, where $\mathbf{a}=(x_g, x_s, t)\in D$ is a vector in the data space (acquisition space), $m(\mathbf{x})$ is function in the model space, where $\mathbf{x}=(x, y, z)\in M$ is a vector in the model space; In the meanwhile $u(\mathbf{a})$ is related to $m(\mathbf{x})$ by a functional $A[m(\mathbf{x})](\mathbf{a})$. Now we define another functional, the data functional: $D[u(\mathbf{a})](\boldsymbol{\tau})$ which is a function of $\boldsymbol{\tau}$. The chain rule of functional differentiation under the linear approximation is

$$\frac{\delta D[m]}{\delta m(\mathbf{x})}=\int\frac{\delta D[A]}{\delta A(\mathbf{a})}\frac{\delta A[m](\mathbf{a})}{\delta m(\mathbf{x})}d\mathbf{a} \tag{B1}$$

The detailed derivation can be found in many text book (e.g. Engel and Dreizler, 2011, Appendix A: Functionals and the functional derivative). If $D[A]$ is an ordinary functional, it reduces to

$$\frac{\delta D[m]}{\delta m(\mathbf{x})}=\frac{\delta D[A]}{\delta A(\mathbf{a})}\frac{\delta A[m]}{\delta m(\mathbf{x})} \tag{B2}$$

In notation of mathematical literature (e.g. Teschl, 1998, Section 8.5; Zhang, 2005, section 1.1.1), the above equation can be written as

$$\frac{\delta(D\circ A)[m]}{\delta m}=\frac{\delta D[A]}{\delta A}\circ\frac{\delta A[m]}{\delta m} \tag{B3}$$

where "$\circ$" signifies the successive functional dependence. In the case of envelope data functional, $E[A](t)\triangleq u^2(t)+u_H^2(t)$, is the envelope operator, and $u_H$ is the Hilbert transform of *u*. Since the envelope operator is applied to a single trace and the envelope is an instantaneous parameter, so (B1) is reduced to

$$\frac{\delta E[m]}{\delta m(\mathbf{x})}=\frac{\delta E[u]}{\delta u}\frac{\delta u[m]}{\delta m(\mathbf{x})}+\frac{\delta E[u_H]}{\delta u_H}\frac{\delta u_H[m]}{\delta m(\mathbf{x})} \tag{B4}$$

From the discussion in Appendix A, we see that the first order functional derivative (or Fréchet derivative) is a linearization of the nonlinear functional dependence of the sensitivity kernel (A5) (we call it *nonlinear functional derivative*). The application of the chain rule (B1) or (B4) is a double linearization which significantly increases the limitation of the applicability of linearizing the nonlinear functional dependence. The functional derivative $\frac{\delta E[u]}{\delta u}$ is only valid when the waveform amplitude (scattered wave amplitude) is much smaller than the incident wave (assume a unit amplitude) so $d(a^2)/da=2a$ can hold. Also as we shown in appendix A, the waveform Fréchet derivative $\frac{\delta u[m]}{\delta m(\mathbf{x})}$ can approximate the nonlinear sensitivity $\left(\frac{\delta u}{\delta m}\right)_{NL}$ only in the case of weak velocity perturbation. For strong nonlinear case, the nonlinear functional is perturbation-dependent. The first order term lost its accuracy and the double linearization makes the situation much worse. In the following we further demonstrate this point in a concise operator form.

In operator form, we write the nonlinear sensitivity operator of envelope to velocity perturbation $\left(\frac{\delta \mathbf{e}}{\delta \mathbf{v}}\right)_{NL}$ (for simplicity we drop the subscript of $\mathbf{e}_W$) in the form

$$\delta \mathbf{e} = \left(\frac{\delta \mathbf{e}}{\delta \mathbf{v}}\right)_{NL} \delta \mathbf{v} \tag{B5}$$

where $\delta \mathbf{e}$ and $\delta \mathbf{v}$ are the variations of envelope data and velocity model, respectively.

In the same way, we can relate $\delta \mathbf{u}$ (waveform data) and $\delta \mathbf{v}$ by a nonlinear derivative operator (sensitivity operator $\mathbf{A}^{(NL)}$),

$$\delta \mathbf{u} = \left(\frac{\delta \mathbf{u}}{\delta \mathbf{v}}\right)_{NL} \delta \mathbf{v} = \mathbf{A}^{(NL)} \delta \mathbf{v} \tag{B6}$$

In this way we can relate $\delta \mathbf{e}$ and $\delta \mathbf{v}$ indirectly through (B5) and (B6):

$$\delta \mathbf{e} = \left(\frac{\delta \mathbf{e}}{\delta \mathbf{u}}\right)_{NL} \left(\frac{\delta \mathbf{u}}{\delta \mathbf{v}}\right)_{NL} \delta \mathbf{v} \tag{B7}$$

This is a generalization of the chain rule to nonlinear functional dependence. Assuming $\mathbf{e}[\mathbf{u}]$ and $\mathbf{u}[\mathbf{v}]$ are both differentiable to arbitrary order, so the nonlinear derivative can be expanded into Taylor series around the points of interests. Then we can write out

$$\left(\frac{\delta \mathbf{e}}{\delta \mathbf{u}}\right)_{NL} = \frac{\delta \mathbf{e}}{\delta \mathbf{u}} + \frac{1}{2!}\frac{\delta^2 \mathbf{e}}{\delta \mathbf{u}^2} \delta \mathbf{u} + \cdots\cdots \tag{B8}$$

Similarly,

$$\left(\frac{\delta \mathbf{u}}{\delta \mathbf{v}}\right)_{NL} = \frac{\delta \mathbf{u}}{\delta \mathbf{v}} + \frac{1}{2!}\frac{\delta^2 \mathbf{u}}{\delta \mathbf{v}^2} \delta \mathbf{v} + \cdots\cdots \tag{B9}$$

We see that the nonlinear functional derivative is function of model perturbation. Stronger the perturbation is, more severe is the nonlinearity.

Substitute (B8) and (B9) into (B7), resulting in

$$\left(\frac{\delta \mathbf{e}}{\delta \mathbf{v}}\right)_{NL} = \frac{\delta \mathbf{e}}{\delta \mathbf{u}}\frac{\delta \mathbf{u}}{\delta \mathbf{v}} + \frac{1}{2!}\left[\frac{\delta \mathbf{e}}{\delta \mathbf{u}}\frac{\delta^2 \mathbf{u}}{\delta \mathbf{v}^2}\delta \mathbf{v} + \frac{\delta^2 \mathbf{e}}{\delta \mathbf{u}^2}\delta \mathbf{u}\frac{\delta \mathbf{u}}{\delta \mathbf{v}}\right] + \cdots\cdots \tag{B10}$$

Take only the first order term, yielding the linearized chain rule of differentiation

$$\frac{\delta \mathbf{e}}{\delta \mathbf{v}} = \frac{\delta \mathbf{e}}{\delta \mathbf{u}}\frac{\delta \mathbf{u}}{\delta \mathbf{v}} \tag{B11}$$

We see that the above relationship is a double linearization which is only valid for case of weak nonlinearity. In the case of strong nonlinearity, the nonlinear functional derivative (sensitivity operator) becomes model dependent and the linearized chain rule is hardly useful.

## Appendix C: Energy scattering and envelope virtual source operator

In this appendix we derive the formulation of energy scattering and the related virtual source operators. Wave equation in operator form can be written as

$$\mathbf{Lu = s}, \quad \mathbf{L_0 u_0 = s}, \tag{C1}$$

where

$$\mathbf{L} = \left[\nabla^2 - \frac{1}{v^2(x)}\frac{\partial^2}{\partial t^2}\right]$$
$$\mathbf{L_0} = \left[\nabla^2 - \frac{1}{v_0^2(x)}\frac{\partial^2}{\partial t^2}\right] \tag{C2}$$

Therefore,

$$\mathbf{L_0 \delta u = -(L - L_0)u = f} \tag{C3}$$

where $\mathbf{f}$ is acting as an equivalent scattering source function caused by the velocity perturbations. In time-space function form it can be written

$$f(x,t) = \left[\frac{1}{v^2(x)} - \frac{1}{v_0^2(x)}\right]\frac{\partial^2}{\partial t^2}u(x,t)$$
$$= \frac{1}{v_0^2(x)}\varepsilon(x)\frac{\partial^2}{\partial t^2}u(x,t) \tag{C4}$$

with $\varepsilon(x)$ as the perturbation index (function) defined by

$$\varepsilon(x) = \frac{v_0^2(x)}{v^2(x)} - 1 = n^2(x) - 1. \tag{C5}$$

In operator form:

$$\delta\mathbf{u} = \mathbf{L_0^{-1} Q \varepsilon} = \mathbf{G_0 Q \varepsilon} \tag{C6}$$

where $\boldsymbol{\varepsilon}$ is the medium perturbation vector (column matrix), $\mathbf{G_0}$ is the background Green's operator and $\mathbf{Q}$ is the virtual source operator (VSO) with kernel $Q(x,x',t)$:

$$Q(x,x',t) = \frac{1}{v_0^2(x)}\frac{\partial^2}{\partial t^2}u(x,t)\delta\left(x - x'\right) \tag{C7}$$

We see that VSO is diagonal. Under weak scattering approximation:

$$\delta u(x,t) \ll u_0(x,t), \tag{C8}$$

and letting $\mathbf{u} \approx \mathbf{u_0}$, leads to

$$\mathbf{Q = Q_0 + \delta Q \simeq Q_0} \tag{C9}$$

with the kernel

$$Q_0(x,x',t) = \frac{1}{c_0^2(x')}\frac{\partial^2}{\partial t^2}u_0(x',t)\delta(x - x') \tag{C10}$$

The above is for perturbation index inversion. For velocity inversion, $\varepsilon(x) \approx -2\delta v(x)/v_0$ for small perturbations, then VSO becomes the familiar form (Pratt et al., 1998)

$$Q(x,x',t) \doteq -\frac{2}{v_0^3(x)}\frac{\partial^2}{\partial t^2}u_0(x,t)\delta\left(x - x'\right) \tag{C11}$$

In frequency-domain, it becomes

$$Q_0(x,x',t)=\frac{2\omega^2}{v_0^3(x')}u_0(x',t)\delta(x-x') \qquad \text{(C12)}$$

## C.1 ***Weak energy scattering for small-scale heterogeneities***

First, we treat the energy scattering for the case of weak scattering with small-scale heterogeneities. Based on(C6), scattered energy can be calculated as

$$\delta\mathbf{e}^2=|\delta\mathbf{u}|^2=\delta\mathbf{u}^*\delta\mathbf{u}=\left(\mathbf{G_0Q\varepsilon}\right)^*\mathbf{G_0Q\varepsilon} \qquad \text{(C13)}$$

The kernel is obtained as a double volume integral,

$$|\delta u|^2(\mathbf{x}_a)=\int_V d^3\mathbf{x}'G_0^*\left(\mathbf{x}_a;\mathbf{x}'\right)Q_0^*\left(\mathbf{x}'\right)\varepsilon\left(\mathbf{x}'\right)\int_V d^3\mathbf{x}''G_0\left(\mathbf{x}_a;\mathbf{x}''\right)Q_0\left(\mathbf{x}''\right)\varepsilon\left(\mathbf{x}''\right), \qquad \text{(C14)}$$

where $\mathbf{x}_a$ is the acquisition location, and $Q_0$ inside the integral is the corresponding virtual source term. Change coordinate system:

$$\xi=\frac{\mathbf{x}'+\mathbf{x}''}{2},\quad \eta=\mathbf{x}'-\mathbf{x}'' \qquad \text{(C15)}$$

Equation (C14) becomes

$$|\delta u|^2(\mathbf{x}_a)=\int_V d^3\xi\int_V d^3\eta G_0^*\left(\mathbf{x}_a;\xi\right)G_0\left(\mathbf{x}_a;\xi+\eta\right)\left[Q_0^*\left(\xi\right)Q_0\left(\xi+\eta\right)\varepsilon\left(\xi\right)\varepsilon\left(\xi+\eta\right)\right] \qquad \text{(C16)}$$

We see that the scattered energy (amplitude square) is dependent on the inner volume integral which in turn depends on the correlation of the perturbation indexes at different points. In case of small-scale, weak heterogeneities, the perturbations are almost random and the correlation radius is very small. Because of the randomness of perturbation, only a small volume with radius $|\eta|<a$, where $a$ is the scale length of the random perturbations, can produce effective scattered energy. Long range interaction between heterogeneities has no effect on the scattering energy due to mutual cancelations. If we treat the perturbation as random function, then (C16) can be written into a formula for random media scattering (Chernov, 1960; Aki and Richards, 1980, vol. II, chapter 13; Wu, 1982; Wu and Aki, 1985; Sato et al., 2012), and the inner volume integral can be taken as correlation integral for random variable

$$|\delta u|^2(\mathbf{x}_a)=\int_V d^3\xi G_0^*\left(\mathbf{x}_a;\xi\right)\int_V d^3\eta G_0\left(\mathbf{x}_a;\xi+\eta\right)\left[Q_0^*\left(\xi\right)Q_0\left(\xi+\eta\right)\tilde{\varepsilon}\left(\xi\right)\tilde{\varepsilon}\left(\xi+\eta\right)\right] \qquad \text{(C17)}$$

where $\tilde{\varepsilon}$ means random variable. Due to the smallness of the effective interaction radius, we can assume the geometric spreading $|G_0\left(\mathbf{x}_a;\xi+\eta\right)|\approx|G_0\left(\mathbf{x}_a;\xi\right)|$ (for homogeneous media, it is simply 1/R), and $|Q_0\left(\xi+\eta\right)|=|Q_0\left(\xi\right)|$. Under these approximations, (C17) can be further reduced to

$$|\delta u|^2(\mathbf{x}_a)\approx\int_V d^3\xi\left|G_0\right|^2\left(\mathbf{x}_a;\xi\right)\left|Q_0\right|^2\left(\xi\right)\varepsilon^2\left(\xi\right)\int_V d^3\eta\frac{G_0\left(\mathbf{x}_a;\eta\right)}{\left|G_0\left(\mathbf{x}_a;\eta\right)\right|}\frac{u_0\left(\eta\right)}{\left|u_0\left(\eta\right)\right|}\tilde{\varepsilon}\left(\xi\right)\tilde{\varepsilon}\left(\xi+\eta\right) \qquad \text{(C18)}$$

In fact $\frac{G_0(\mathbf{x}_a;\eta)}{\left|G_0(\mathbf{x}_a;\eta)\right|}$ and $\frac{u_0(\eta)}{\left|u_0(\eta)\right|}$ are local phase functions and can be expressed as

$$\frac{G_0(\mathbf{x}_a;\eta)}{\left|G_0(\mathbf{x}_a;\eta)\right|}=\exp[i\mathbf{k}_s\cdot\eta],\quad \frac{u_0(\eta)}{\left|u_0(\eta)\right|}=\exp[i\mathbf{k}_i\cdot\eta] \tag{C19}$$

where $\mathbf{k}_s$ and $\mathbf{k}_i$ are scattering and incident wavenumbers (in frequency domain), respectively. So the inner volume integral of (C18) can be taken as ensemble average and the integral represents the power spectra of the random heterogeneities. In the extreme case, i.e. when the small-scale velocity perturbations become uncorrelated, (C18) becomes:

$$|\delta u|^2(\mathbf{x}_a)\approx\int_V d^3\xi\left|G_0\right|^2(\mathbf{x}_a;\xi)\left|Q_0\right|^2(\xi)\,\varepsilon^2(\xi). \tag{C20}$$

This implies that the total power of the scattered waves from all the small scatterers is just the sum of scattered powers from individual scatterers. This is the Born energy scattering. In frequency-domain, the above equation can be written explicitly as

$$\delta e^2(\mathbf{x}_a)=|\delta u|^2(\mathbf{x}_a)\approx\int_V d^3\xi\left|G_0\right|^2(\mathbf{x}_a;\xi)\left[\frac{\omega^4}{v_0^4}\left|u_0\right|^2(\xi)\right]\varepsilon^2(\xi), \tag{C21}$$

and in time-domain,

$$\delta e^2(\mathbf{x}_a)=|\delta u|^2(\mathbf{x}_a)\approx-\int_V d^3\xi\left|G_0\right|^2(\mathbf{x}_a;\xi)*\frac{1}{v_0^4(x')}\frac{\partial^4}{\partial t^4}\left|u_0\right|^2(\xi)\,\varepsilon^2(\xi) \tag{C22}$$

This is the *Born scattering* for the energy case. Scattered energy from each perturbation is independent of others and the total scattered energy is just the sum of those from individual scatterers. It can be seen that in frequency-domain the energy scattering is proportional to the 4$^{th}$ power of frequency, and in time domain it corresponds to a 4$^{th}$ order of differentiation with time. We can take the square-root of energy as envelope data. However, the linearity of energy scattering will be somehow lost. Based on (C21) we derive the virtual source operator for the Born energy scattering: For weak scattering we further make approximation about the perturbation index $\varepsilon$ and $\varepsilon^2$. Since the velocity perturbation is assumed much small that the background velocity, so

$$\varepsilon^2=(n^2-1)^2=(n-1)^2(n+1)^2\approx4\left(\frac{\delta v}{v_0}\right)^2 \tag{C23}$$

based on the approximations:

$$\begin{aligned}n-1&=\frac{v_0}{v}-1=\frac{v_0-v}{v}=\frac{-\delta v}{v}\approx\frac{-\delta v}{v_0},\\ n+1&=\frac{v_0}{v}+1=\frac{v_0+v}{v}\approx2\end{aligned} \tag{C24}$$

In fact, from (C22) we see that the scattering energy is proportional to the square of slowness perturbation. The inversion could be set up for the perturbation index or equivalently for the squared slowness. However, under the small perturbation assumption, it makes no differences to set up for velocity inversion. Based on (C22) and (C23), the kernel of VSO for velocity perturbation is derived as

$$Q_v^{(e)}(x,x',t) = \left| \frac{4}{v_0^6(\mathbf{x})} \frac{\partial^4}{\partial t^4} \left| G_0 \right|^2 (x',t;x_s) \delta\left(x - x'\right) \right. \tag{C25}$$

Note that the VSO is related to the squared perturbation $(\delta v)^2$, such that

$$\delta \mathbf{e}^2 = \left| \delta \mathbf{u} \right|^2 = \left| \mathbf{G_0} \right|^2 \mathbf{Q}_v^{(e)} (\delta \mathbf{v})^2 = \mathbf{G}_0^{(e)} \mathbf{Q}_v^{(e)} (\delta \mathbf{v})^2 \tag{C26}$$

**C.2 *Strong energy scattering: Boundary reflections***

In case of strong scattering, such as boundary reflections, the wave and energy scattering can be quite different from the weak scattering. Boundary reflection is formed by coherent scattering by large volume strong-contrast inclusions (such as salt bodies, carbonate rocks, etc.). From (C18) we may also see that when the correlation function $N(\eta)$ become strong, the cross-terms become important, so the mutual interference may not be ignored. In case of large homogeneous inclusion, the cross-terms can be of the same strength of the auto-correlation term. Such as in the case of salt body, the mutual interference of scattered waves from all the volume elements leads to the total cancellation of scattered waves from the salt interior and the formation of the boundary reflections. This is why boundary reflection is treated totally different mathematically from volume heterogeneities. Wave equations are set for inner and outer regions and boundary conditions are to be matched across boundaries. In principle, Lippmann-Schwinger equation should be able to treat all cases of wave scattering in heterogeneities media. However, the connection between volume scattering and boundary reflection has not been well established based on our knowledge. Some authors tried to use the renormalization theory to build a bridge connecting the two treatments (e.g. Kirkinis, 2008; Wu et al., 2016) but only partly successful. In this paper we will use the traditional boundary reflection calculation to derive the virtual source operator for envelope inversion.

Scattered waves from boundary scattering can be modeled as a summation of reflected waves from all the sharp boundaries inside the considered volume. Assume $\Omega$ as the collection of boundaries surrounding different domains, $\Gamma \equiv \{\Gamma_n\}$, $n = 1, 2, \cdots N$. Each domain has smooth velocity variation inside the domain. For any curved boundary (interface) $\Gamma_1$, the scattered waves can be formulated by a representation integral (Helmholtz-Kirchhoff integral, we will simply call Kirchhoff integral),

$$\delta u(\mathbf{x}) = -\int_{\Gamma_1} \left\{ u_s(\mathbf{x}_0) \frac{\partial}{\partial n} G_F(\mathbf{x};\mathbf{x}_0) - G_F(\mathbf{x};\mathbf{x}_0) \frac{\partial}{\partial n} u_s(\mathbf{x}_0) \right\} ds(\mathbf{x}_0) \tag{C27}$$

where $G_F$ is forward-scattering approximated Green's function in the exterior region (for smooth exterior media, $G_F = G$), $ds(\mathbf{x}_0)$ is a boundary element at $\mathbf{x}_0$ on the boundary $\Gamma_1$, $u_s(\mathbf{x}_0)$ is the scattered field at that point, $\hat{n}$ is the outward normal and $\partial / \partial n$ is the normal derivative at that point. $u_s(\mathbf{x}_0)$ can be obtained by solving the boundary integral equation (BIE). However, for inversion purpose, we need to obtain the explicit relation with the local velocity contrast. Here we invoke the P.O. (physical optics) approximation or Kirchhoff approximation (Birkhout, 1987) for smooth boundaries. In P.O. approximation, the reflection is calculated by short-wavelength approximation (plane wave incident on a plane boundary – the tangent plane), but the propagation is implemented by wave theory). Because of the smoothness of the boundary, scattered wavefield is approximated by the product of incident wave and local reflection coefficient $\gamma(\mathbf{x}_0,\hat{n},\theta)$ which is reflection-angle dependent. In this paper, we further simplify the relation to the case of small reflection angles, so $\gamma(\mathbf{x}_0,\hat{n})$ is directly proportional to the local impedance contrast, in our case of constant density, to the local velocity contrast. Therefore, the scattered field $u_s(\mathbf{x}_0)$ on the boundary can be approximated as

$$u_s(\mathbf{x}_0)=\gamma(\mathbf{x}_0,\hat{n})u_0(\mathbf{x}_0)=\gamma(\mathbf{x}_0,\hat{n})G_0(\mathbf{x}_0,\mathbf{x}_s) \tag{C28}$$

where $u_0(\mathbf{x}_0)=G_0(\mathbf{x}_0,\mathbf{x}_s)$ is the local incident field on the boundary. Substitute (C28) into (C27), yielding

$$\delta u(\mathbf{x}_g)=-\left\{\int_{\Gamma_1}\gamma(\mathbf{x}_0,\hat{n})\left[G_F(\mathbf{x}_0;\mathbf{x}_s)\frac{\partial}{\partial n}G_F(\mathbf{x}_g;\mathbf{x}_0)-G_F(\mathbf{x}_g;\mathbf{x}_0)\frac{\partial}{\partial n}G_F(\mathbf{x}_0;\mathbf{x}_s)\right]ds(\mathbf{x}_0)\right\} \tag{C29}$$

Due to the fact that $\frac{\partial}{\partial n}G_F(\mathbf{x}_0;\mathbf{x}_s)\doteq-\frac{\partial}{\partial n}G_F(\mathbf{x}_g;\mathbf{x}_0)$ for small-angle reflections, above equation can be simplified to

$$\begin{aligned}\delta u(\mathbf{x}_g,\mathbf{x}_s)&=-2\int_{\Gamma_1}\gamma(\mathbf{x}_0,\hat{n})\frac{\partial}{\partial n}G_F(\mathbf{x}_g;\mathbf{x}_0)G_F(\mathbf{x}_0;\mathbf{x}_s)ds(\mathbf{x}_0)\\&=2\int_{\Gamma_1}\gamma(\mathbf{x}_0,\hat{n})G_F(\mathbf{x}_g;\mathbf{x}_0)\frac{\partial}{\partial n}G_F(\mathbf{x}_0;\mathbf{x}_s)ds(\mathbf{x}_0)\end{aligned} \tag{C30}$$

In order to be consistent with the volume integral formulation, we may extend the surface integral into volume integral by assuming zero-reflection outside all the real boundaries $\Gamma\equiv\{\Gamma_n\}$, resulting in

$$\delta u(\mathbf{x}_g,\mathbf{x}_s)=2\int_V\gamma_\Gamma(\mathbf{x}_0,\hat{n})_\Gamma G_F(\mathbf{x}_g;\mathbf{x}_0)G_n(\mathbf{x}_0;\mathbf{x}_s)ds(\mathbf{x}_0) \tag{C31}$$

where $G_n=\partial G_F/\partial n$ and $\gamma_\Gamma(\mathbf{x}_0,\hat{n})$ exists only on the boundary $\Gamma$, defined as

$$\gamma_\Gamma(\mathbf{x},\hat{n})\triangleq\gamma(\mathbf{x},\hat{n})\delta(\mathbf{x}-\mathbf{x}_0),\quad \mathbf{x}\in V;\mathbf{x}_0\in\Gamma \tag{C32}$$

Now we derive the energy scattering for this case. Since we are dealing with energy scattering and neglecting the interference between surface elements, so the energy scattering of boundary elements can be treated in the same way as for volume elements. Taking small-angle approximation, we can replace $G_n$ by $G_F$ ($G_n ds=G_F\cos\theta ds\doteq G_F ds$). Similar to the derivation of (C20), we obtain the energy scattering for the boundary reflection case,

$$|\delta u|^2(\mathbf{x}_g) \approx 4\int_V d^3\xi \left|G_F\right|^2(\mathbf{x}_g;\xi)\left|G_F\right|^2(\xi;\mathbf{x}_s)\gamma_\Gamma^2(\xi,\hat{n}) \\ = 4\int_V d^3\mathbf{x}\left|G_F\right|^2(\mathbf{x}_g;\mathbf{x})\left|G_F\right|^2(\mathbf{x};\mathbf{x}_s)\gamma_\Gamma^2(\mathbf{x},\hat{n}) \tag{C33}$$

Note that in the derivation, we assumed the independence of $\gamma_\Gamma(\mathbf{x}_0,\hat{n})$ at each point. However, when we use the P.O. approximation, the reflection coefficient is calculated as if the local boundary can be approximated by an infinite plan (tangent approximation), which is equivalent to total correlated scattering between neighboring points (boundary is locally smooth to the dominant frequency). Only when summing up scattered energy at the receiver, we consider the reflected energy from each boundary element as independent from each other (boundary is irregular for the whole reflection records).

In order to compare with instantaneous amplitude records obtained by applying envelope operator to reflection seismograms, we need to revoke the *sparse reflection approximation*, which implies narrow-pulse for relatively smooth boundary. For narrow-pulse, in the limiting case, a δ-pulse, the reflected signals from different surface elements will not interfere so they become separate arrivals in time-domain. Then (C33) can be expressed as

$$|\delta u|^2(\mathbf{x}_g,t) \approx 4\int_V d^3\mathbf{x}\left|G_F\right|^2(\mathbf{x}_g;\mathbf{x},t) * \left|G_F\right|^2(\mathbf{x};\mathbf{x}_s,t)\gamma_\Gamma^2(\mathbf{x},\hat{n}) \\ = 4\int_V d^3\mathbf{x}\left|G_F\right|^2(\mathbf{x}_g;\mathbf{x},t) * \left|G_F\right|^2(\mathbf{x};\mathbf{x}_s,t)\gamma_\Gamma^2(\mathbf{x},\hat{n}), \quad \mathbf{x}\in\delta\Gamma(t) \tag{C34}$$

where “*” stands for time-convolution, and $\delta\Gamma(t)$ means on the boundary surface elements which are intersected by the source and receiver Green's functions at time t so that they can contribute to the integration.

To obtain the VSO for this case, we need to write out the relation of $\gamma(\mathbf{x}_0,\hat{n})$ with velocity perturbations. For acoustic media, reflection coefficient by a plane boundary is

$$\gamma = \frac{Z_2^{(n)} - Z_1^{(n)}}{Z_2^{(n)} + Z_1^{(n)}} \tag{C35}$$

where $Z^{(n)}$ is the normal component of acoustic impedance

$$Z^{(n)} = \rho c / \cos\theta \tag{C36}$$

where $\theta$ is the reflection angel. With constant density and small angle reflection ($\cos\theta \simeq 1$), we have

$$\gamma \approx \frac{v_2 - v_1}{v_2 + v_1} = \frac{\Delta v}{2v_0 + \Delta v} \tag{C37}$$

where $v_0 = v_1$ and $\Delta v = v_2 - v_1$. Then the energy reflection coefficient is

$$\gamma_e = \gamma^2 \approx \frac{(v_2 - v_1)^2}{(v_2 + v_1)^2} = \frac{(\Delta v)^2}{(2v_0 + \Delta v)^2} \tag{C38}$$

From (C33) and (C38) we derive the VSO for energy reflection coefficient,

$$\left|Q_\gamma\right|^2 (x, x', t) \doteq 4\left|G_F\right|^2 (x', t; x_s)\delta\left(x - x'\right) \tag{C39}$$

and for amplitude reflection,

$$\left|Q_\gamma\right| (x, x', t) \doteq 2\left|G_F\right| (x', t; x_s)\delta\left(x - x'\right) \tag{C40}$$

In operator form, (C34) can be written respectively as

$$\delta\mathbf{e}^2 = |\,\delta\mathbf{u}\,|^2 = \left|\mathbf{G}_F\right|^2 \left|\mathbf{Q}_\gamma\right|^2 \boldsymbol{\gamma}^2 \tag{C41}$$

Incorporating (C38) into above equation, finally we obtain

$$\delta\mathbf{e}^2 = |\,\delta\mathbf{u}\,|^2 = \left|\mathbf{G}_F\right|^2 \left|\mathbf{Q}_v\right|^2 \delta\mathbf{v}^2 \tag{C42}$$

where $\left|\mathbf{Q}_v\right|^2$ is the VSO for velocity perturbation with the kernel

$$\left|Q_v\right|^2 (x, x', t) \doteq \frac{4}{(2v_0 + \Delta v)^2}\left|G_F\right|^2 (x', t; x_s)\delta\left(x - x'\right) \tag{C43}$$

**References:**

Aki and Richards, (1980), "Quantitative Seismology: Theory and Methods", vol. II, W.H.Freeman and Company.

Aki and Richards, (2002), "Quantitative Seismology", second edition, University Science Books.

Baeten, G., de Maag, J.W., Plessix, R.-E., Klaassen, M., Qureshi, T., Kleemeyer, M., ten Kroode, F. & Zhang, R., (2013). The use of the low frequencies in a full waveform inversion and impedance inversion land seismic case study, Geophys. Prospect., 61, 701–711.

Bharadwaj, P, Mulder, W. & Drijkoningen, G. (2016). Full waveform inversion with an auxiliary bump functional. Geophys. J. Int., 206, 1076–1092.

Birkhout, A.J., (1987), "Applied Seismic Wave Theory", Elsevier.

Bozdag, E., J. Trampert, & J. Tromp, (2011), Misfit functions for full wave-form inversion based on instantaneous phase and envelope measurements, Geophys. J. Int., 185, 845–870.

Bunks, C., Saleck F.M., Zaleski S., & Chavent G., (1995). Multiscale seismic waveform inversion: Geophysics, 60, 1457–1473.

Chen, G.X., R.S. Wu and S.C. Chen., 2017. The nonlinear data functional and multiscale seismic envelope inversion: Algorithm and methodology for application to salt structure inversion. SEG Technical Program Expanded Abstracts 2017: pp. 1697-1701.

Chen, G.X., R.S. Wu and S.C. Chen., 2018a. Reflection Multi-scale Envelope Inversion. Geophysical Prospecting. Accepted. DOI: 10.1111/1365-2478.12624.

Chen, G.X., R.S. Wu , Wang Y and S.C. Chen., 2018b. Multi-scale signed envelope inversion. Journal of Applied Geophysics, 153：113-126.

Chernov, L. A., 1960, Wave Propagation in a Random Medium, McGraw-Hill, New York, 1960.

Chi, B., Dong, L. & Liu, Y., (2014). Full waveform inversion method using envelope objective function without low frequency data, J. Appl. Geophys.,109, 36–46.

Choi Y, Alkhalifah T., 2013, Frequency-domain waveform inversion using the phase derivative, *Geophysical Journal International*, **195**(3):1904-1916.

Engel, E., R.M. Dreizler, (2011), Appendix A: Functional and the functional derivative. In: E.Engel, R.M. Dreizler, Density Functional Theory, Theoretical and Mathematical Physics, pp. 403–531. DOI 10.1007/978-3-642-14090-7.

Fichtner A, Kennett B L N, Igel H, 2008, Theoretical background for continental- and global-scale full-waveform inversion in the time-frequency domain, *Geophysical Journal International*, **175**(2):665-685.

Kirkinis, E., (2008), Renormalization group interpretation of the Born and Rytov approximations, J. Opt. Soc. Am, A., 25, 2499-2508.

Lailly, P., (1983). The seismic inverse problem as a sequence of before stack migration, in Proc. of Conf. on Inverse Scattering, Theory and Applications, SIAM, Philadelphia, Pennsylvania, 206–220.

Luo, Y. and Schuster, G. T., 1991, Wave-equation traveltime inversion, *Geophysics*, **56**(5): 645-653.

Luo, J. & Wu, R.S., (2015), Seismic envelope inversion: Reduction of local minima and noise resistance, Geophys. Prospect., 63, 597-614.

Luo, J., Wu R.S and Gao F., 2016, Time-domain full-waveform inversion using instantaneous phase with damping, SEG Technical Program Expanded, 1472-1476.

Mora, P., (1989). Inversion= migration+tomography, Geophysics, 54, 1575-1586.

Plessix, R., G. Baeten, J. W. de Maag, M. Klaassen, R. Zhang & Z. Tao, (2010), Application of acoustic full waveform inversion to a low-frequency large-offset land data set, in roceedings of the SEG Technical Program, Expanded Abstracts, pp: 930-934.

Pratt, R. G., C. Shin, and G. J. Hicks, (1998), Gauss-Newton and full Newton methods in frequency-space seismic waveform inversion: Geophys. J. Int., 133, 341–362.

Sato, H., M.C. Fehler and T. Maeda, (2012), "Seismic wave propagation and scattering in the heterogeneities earth", second edition, Springer.

Shin, C. & Cha Y. H., (2008), Waveform inversion in the Laplace domain, *Geophys. J. Int*., 173, 922–931.

Shin C. & Cha, Y. H., (2009), Waveform inversion in the Laplace-Fourier domain, *Geophys. J. Int.*, 177, 1067–1079.

Tarantola, A., 1984a, Linearized inversion of seismic reflection data, *Geophys. Prospecting*, 32, 998-1015.

Tarantola, A., 1984b, Inversion of seismic reflection data in the acoustic approximation: Geophysics, **49**, 1259–1266.

Tarantola, A., (2005), "Inverse Problem Theory and Methods for Model Parameter Estimation", SIAM.

Teschl, G., (1998). "*Nonlinear Functional Analysis"*, University of Vienna. Available at: http://www.mat.univie.ac.at/~gerald/ftp/book-nlfa/

Tromp, J., Tape, C. & Liu, Q., 2005, Seismic tomography, adjoint methods, time reversal, and banana-donut kernels, *Geophys. J. Int.*, **160**, 195–216.

Vigh, D., J. Kapoor, N. Moldoveanu, & H. Li, (2011), Breakthrough acquisition andtechnologies for subsalt imaging: Geophysics, 76, WB41-WB51.

Virieux, J. & Operto, S., (2009). An overview of full waveform inversion in exploration geophysics, Geophysics, 74(6), WCC127–WCC152.

Wu, R.S., (1982a), Attenuation of short period seismic waves due to scattering, *Geophys. Res. Lett.*, **9,** 9-12.

Wu, R.S. ,(1985), Multiple scattering and energy transfer of seismic waves, -separation of scattering effect from intrinsic attenuation I. Theoretical modelling, *Geophys. J. R. astr. Soc.*, **82,** 57-80.

Wu, R.S., and K. Aki, (1985), Elastic wave scattering by a random medium and the small scale inhomogeneities in the lithosphere, *J. Geophys. Res.*, **90** 10261-10273.

Wu, R.S & Chen, G. X. (2017a). Multi-scale seismic envelope inversion for strong-nonlinear FWI, 2017 CGS/SEG International Geophysical Conference, Qingdao, China, 17-20.

Wu, R.S & Chen, G. X. (2017b). New Fréchet derivative for envelope data and multi-scale envelope inversion, in 79th EAGE conference & Exhibition 2017.

Wu, R.S, Luo, J.R. & Chen, G. X. (2016). Seismic envelope inversion and renormalization group theory: Nonlinear scale separation and slow dynamics, in Proceedings of the SEG Technical Program, Expanded Abstracts. pp 1346-1351.

Wu, R.S., J. Luo & B. Wu, (2013), Ultra-low-frequency information in seismic data and envelope inversion: in Proceedings of the SEG Technical Program, Expanded Abstracts. pp. 3078-3082.

Wu, R.S., Luo, J. and Wu, B., (2014), Seismic envelope inversion and modulation signal model, *Geophysics*, 79, WA13-WA24.

Wu, R.S. & Zheng, Y., (2014), Nonlinear partialderivative and its De Wolf approximation for nonlinear seismic inversion, *Geophys. J. Int*., 196, 1827-1843.

Wu, R.S., B. Wang & C. Hu, (2015), Inverse thin-slab propagator in T-matrix formalism based on nonlinear sensitivity kernel for wave-equation tomography, *Inverse Problems*, 31, 115004.

Yan, R., Guan, H., Xie, X.B. & Wu, R.S., (2014), Acquisition aperture correction in angle-domain towards the true-reflection reverse time migration, *Geophysics*, 79, S241-S250.

Zeng, Y., F. Su and K. Aki, (1991), Scattering wave energy propagation in a random isotropic scattering medium 1. Theory, *J. Geophys. Research*., 96, 607-619.

Zhang, K.C., (2005). *Methods in Nonlinear Analysis*, Springer-Verla